\PassOptionsToPackage{pagebackref=true}{hyperref}
\documentclass[sigconf]{acmart}%
\pdfoutput=1

\usepackage{booktabs} 
\usepackage{multirow}
\usepackage{makecell}

\setcopyright{rightsretained}

\setcitestyle{square}

\usepackage{wrapfig}

\usepackage{soul}
\usepackage{color}

\definecolor{todoCol}{rgb}{1,0.5,0}
\definecolor{blueColor}{rgb}{0,0.5,1}

\definecolor{redColor}{rgb}{1,0,0}
\definecolor{darkRedColor}{rgb}{0.5,0,0}

\begin{document}

\title{3D Sketching using Multi-View Deep Volumetric Prediction}

\author{Johanna Delanoy}
\affiliation{%
	\institution{Inria Universit\'e C\^ote d'Azur}}
\author{Mathieu Aubry}
\affiliation{%
	\institution{LIGM (UMR 8049), Ecole des Ponts}}
\author{Phillip Isola}
\affiliation{%
	\institution{OpenAI}}
\author{Alexei A. Efros}
\affiliation{%
	\institution{UC Berkeley}}
\author{Adrien Bousseau}
\affiliation{%
	\institution{Inria Universit\'e C\^ote d'Azur}}
\renewcommand{\shortauthors}{Delanoy, Aubry, Isola, Efros and Bousseau}


\begin{teaserfigure}
   \includegraphics[width=\textwidth]{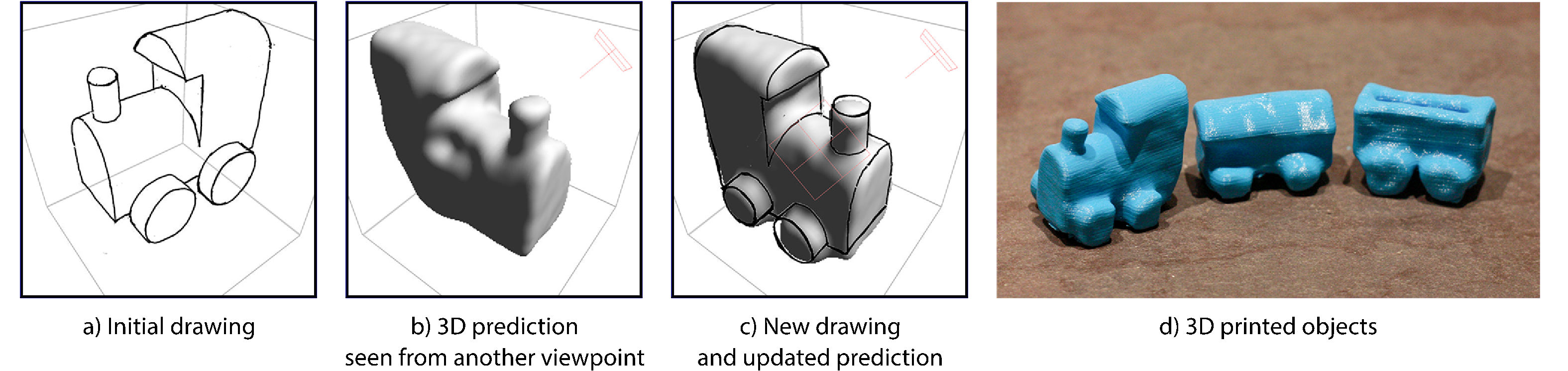}
   \caption{Our sketch-based modeling system can process as little as a single perspective drawing (a) to predict a
   volumetric object (b). Users can refine this prediction and complete it with novel parts by providing additional drawings from other viewpoints (c). This iterative sketching workflow allows quick 3D concept exploration and rapid prototyping (d). }
   \label{fig:teaser}
 \end{teaserfigure}

\begin{abstract}
Sketch-based modeling strives to bring the ease and immediacy of drawing to the 3D world. However, while drawings are easy for humans to create, they are very challenging for computers to interpret due to their sparsity and ambiguity. 
We propose a data-driven approach 
that tackles this challenge by learning to reconstruct 3D shapes from one or more drawings. At the core of our approach is a deep convolutional neural network (CNN) that predicts occupancy of a voxel grid from a line drawing. This CNN provides an initial 3D reconstruction as soon as the user completes a single drawing of the desired shape. We complement this single-view network with an \emph{updater CNN} that refines an existing prediction given a new drawing of the shape created from a novel viewpoint. A key advantage of our approach is that we can apply the updater iteratively to fuse information from an arbitrary number of viewpoints, without requiring explicit stroke correspondences between the drawings. We train both CNNs by rendering synthetic contour drawings from hand-modeled shape collections as well as from procedurally-generated abstract shapes.
Finally, we integrate our CNNs in an interactive modeling system that allows users to seamlessly draw an object, rotate it to see its 3D reconstruction, and refine it by re-drawing from another vantage point using the 3D reconstruction as guidance.

\textcolor{darkRedColor}{ This is the authors version of the work. It is posted by permission of ACM for your personal use. Not for redistribution. The definite version will be published in PACMCGIT.}
\end{abstract}

%
%




\maketitle

\section{Introduction}
The ambition of sketch-based modeling is to bring the ease and immediacy of sketches to the 3D world to provide \emph{``an environment for rapidly conceptualizing and editing approximate 3D scenes''} \cite{Zeleznik1996}. However, while humans are extremely good at perceiving 3D objects from line drawings, this task remains very challenging for computers. 
In addition to the ill-posed nature of 3D reconstruction from a 2D input, line drawings lack important shape cues like texture and shading, are often composed of approximate sketchy lines, and even when multiple drawings of a shape are available, their level of inaccuracy prevents the use of geometric algorithms like multi-view stereo. We introduce a data-driven sketch-based modeling system that addresses these challenges by \emph{learning} to predict 3D volumes from one or several freehand bitmap drawings. Our approach builds on the emerging field of generative deep networks, which recently made impressive progress on image \cite{Chen2017} and shape synthesis \cite{Fan_cvpr2017} but has been little used for interactive creative tasks.

Figure~\ref{fig:teaser} illustrates a typical modeling session with our system. The user starts by drawing an object from a 3/4 view, which is the viewpoint preferred by designers to illustrate multiple sides of a shape in a single drawing. Thanks to training on a large collection of 3D shapes, our approach produces a complete volumetric reconstruction of the object, including occluded parts. This initial reconstruction allows the user to rotate the object and inspect it from a different vantage point.
The user can then either re-draw the object from this new viewpoint to correct errors in the reconstruction, or move on to drawing new parts of the object. In both cases, the temporary 3D reconstruction acts as a reference that significantly helps the user create new drawings of the 3D shape. Since all interactions occur in a shared 3D space, this workflow provides us with multiple registered drawings of the object along with their respective calibrated cameras, which form the input to our 3D reconstruction algorithm.



At the core of our system are deep convolutional neural networks (CNNs) that we train to predict occupancy in a voxel grid, given one or several contour drawings as input. These CNNs form a flexible and robust 3D reconstruction engine that can interpret bitmap drawings without requiring complex, hand-crafted optimizations \cite{lipson1996optimization,Xu2014} nor explicit correspondences between strokes in multiple views \cite{bae2008ilovesketch,Rivers2010}.
However, applying deep learning to sketch-based modeling raises several major new challenges. First, 
we need a network architecture capable of fusing the information provided by multiple, possibly inconsistent, drawings. Our solution combines a single-view network, which generates an initial reconstruction from one drawing, with an \emph{updater network} that iteratively refines the prediction as additional drawings are provided.
This iterative strategy allows us to handle drawings created from an arbitrary number of views, achieving a continuum between single-view \cite{Gingold09} and multi-view \cite{bae2008ilovesketch} sketch-based modeling systems. 

The second challenge we face is access to training data, as collecting thousands of hand-drawings registered with 3D objects would be very costly and time consuming. Similarly to prior data-driven approaches \cite{eitz2012sbsr,xie2013sketch,huang2016shape,gen2016interactive}, we alleviate the need for collecting real-world drawings by generating \emph{synthetic} line drawings from 3D objects using non-photorealistic rendering. This allows us to easily adapt our system to the design of different types of objects by generating training data that is representative of such objects.
We first illustrate this capability by training an instance of our system with a dataset of chairs, and another instance with a dataset of vases. We then target the design of more general objects by training our system with abstract shapes assembled from simple geometric primitives (cuboids, cylinders). We used this latter instance to model a variety of man-made objects and show that it generalizes well to unseen object categories. Finally, we describe how to co-design the training data and the user interface to reduce ambiguity in the prediction. In particular, we restrict viewpoints for the first drawing to avoid depth ambiguity for the single-view network, while we allow greater freedom for the subsequent drawings that are handled by the updater network.

Once trained, our system can generate a coherent multi-view prediction in less than a second, which makes it suited for interactive modeling. One restriction of our current implementation is that the resolution of the voxel grid hinders the recovery of thin structures. We thus target quick 3D design exploration rather than detailed modeling.

In summary, we introduce an interactive sketch-based modeling system capable of reconstructing a 3D shape from one or several freehand bitmap drawings. In addition to the overall system, we make the following technical contributions\footnote{Networks and databases are available online at https://ns.inria.fr/d3/3DSketching/}:
\vspace{-0.5mm}
\begin{itemize}
\item An iterative \emph{updater network} that predicts coherent 3D volumes from multiple drawings created from different viewpoints .
\item A multi-view drawing interface that we co-design with our synthetic data to help users create drawings similar to the ones used for training.
\end{itemize}
Note that our approach is modular and could adapt to other drawing techniques and shapes than the ones used in this paper.

\section{Related work}
Our work builds on recent progress in deep learning to tackle the long standing problem of sketch-based modeling. We refer the interested reader to recent surveys for extended discussions of these two fields \cite{Srinivas16,OSSJ09,Cordier2016}.

\subsection{Sketch-based modeling}
The problem of creating 3D models from line drawings has been an active research topic in computer graphics for more than two decades \cite{Zeleznik1996,lipson1996optimization,igarashi1999teddy}. While sketching is one of the most direct ways for people to represent imaginary 3D objects, recovering 3D information from 2D strokes poses significant challenges since an infinity of 3D shapes can potentially re-project on the same drawing \cite{barrow81}. Various approaches have been proposed to tackle the inherent ambiguity of this inverse problem.

\emph{Constrained-based approaches} assume that the lines in a drawing represent specific shape features, from which geometric constraints can be derived and imposed in an optimization framework. Popular constraints include surface orientation along smooth silhouettes \cite{Malik89}, orthogonality and parallelism of edges on polyhedral models \cite{lipson1996optimization}, symmetry \cite{Cordier2013}, and surface developability \cite{jung2015} among others. However, the assumptions made by these methods often restrict them to specific classes of shapes, or specific drawing techniques such as polyhedral scaffolds \cite{schmidt2009analytic}, curvature-aligned cross-sections \cite{shao2012crossshade,Xu2014} or cartoon isophotes \cite{Xu2015}. In addition, most of these methods require clean vector drawings as input to facilitate the detection of suitable constraints, as well as to compute the various energy terms that drive the optimization. Unfortunately, converting rough sketches into clean vector drawings is a difficult problem in its own right \cite{FLB16}, while methods capable of directly recovering 3D information from noisy drawings are prohibitively expensive \cite{IBB15}. In this work, we bypass all the challenges of defining, detecting and optimizing for multiple geometric constraints by training a deep convolutional neural network (CNN) to automatically predict 3D information from bitmap line drawings.

\emph{Interactive approaches} reduce ambiguity in 3D reconstruction by leveraging user annotations. Single-image methods allow users to create 3D models from existing imagery by snapping parametric shapes to image contours \cite{Chen2013,Shtof2013} or by indicating geometric constraints such as equal length and angle, alignment and symmetry \cite{Gingold09} or depth ordering \cite{Sykora14-TOG}. Other methods adopt an incremental workflow where users progressively build complex shapes by drawing, modifying and combining simple, easy to reconstruct 3D parts. Existing systems differ in the type of assumptions they make to reconstruct intermediate shapes from user strokes, such as smooth shapes inflated from silhouettes \cite{igarashi1999teddy,Nealen07}, symmetric or multi-view pairs of 3D curves related by epipolar constraints \cite{Orbay2012,bae2008ilovesketch}, curves lying on pre-defined planes or existing surfaces \cite{bae2008ilovesketch,Zheng16}, visual hulls carved from orthogonal viewpoints \cite{Rivers2010}. The main drawback of such methods is that users have to mentally decompose the shape they wish to obtain, and construct it by following a carefully ordered series of sketching operations, often performed from multiple viewpoints. In contrast, while our system supports incremental modeling, our CNN-based reconstruction engine does not rely on restrictive assumptions on the drawn shapes and allows users to draw a complete object from one viewpoint before visualizing and refining it from other viewpoints.

\emph{Data-driven approaches} exploit large collections of 3D objects to build priors on the shapes that users may draw. Early work focused on retrieving complete objects from a database \cite{Funkhouser2003,eitz2012sbsr}, which was later extended to part-based retrieval and assembly \cite{Lee2008,xie2013sketch} and to parameter estimation of pre-defined procedural shapes \cite{gen2016interactive,huang2016shape}. While our approach also learns shape features from object databases, we do not require these objects to be expressible by a known parametric model, nor be aligned and co-segmented into reconfigurable parts. Instead, our deep network learns to generate shapes directly from pairs of line drawings and voxel grids, which allows us to train our system using both existing 3D model databases and procedurally-generated shapes. Our approach is also related to the seminal work of Lipson and Shpitalni \shortcite{Lipson2000}, who used a database of random polyhedrons to learn geometric correlations between 2D lines in a drawing and their 3D counterpart. The considered correlations include the angles between pairs and triplets of lines, as well as length ratios. These priors are then used to evaluate the quality of a 3D reconstruction in a stochastic optimization. In a similar spirit, Cole et al.~\shortcite{cole2012shapecollage} generate a large number of abstract blobs to serve as exemplars for a patch-based synthesis algorithm that converts line drawings into normal maps. While we build on these initial attempts, deep learning alleviates the need for custom feature extraction and optimization and allows us to handle a wider diversity of shapes. In addition, we integrate our 3D reconstruction engine in an interactive system capable of fusing information from multiple sketches drawn from different viewpoints.

\subsection{Deep learning}
Our work is motivated by the recent success of deep convolutional neural networks in solving difficult computer vision problems such as image classification \cite{krizhevsky2012imagenet}, semantic segmentation \cite{long_shelhamer_fcn}, depth and normal prediction \cite{eigen2015predicting,wang2015designing}. In particular, our single-view volumetric reconstruction network follows a similar encoder-decoder architecture as depth prediction networks, although we also propose a multi-view extension that iteratively refines the prediction as new sketches are drawn by the user. This extension is inspired by iterative networks that implement a feedback loop to impose structural constraints on a prediction, for instance to refine hand \cite{oberweger2015training} and human pose \cite{carreira2015human}.

Our architecture also shares similarities with deep networks tailored to multi-view 3D reconstruction. Choy et al.~\shortcite{choy20163d} train a recurrent neural network (RNN) to predict a voxel reconstruction of an object from multiple uncalibrated photographs. Similarly, our iterative updater network can be seen as a recurrent network that is unrolled to simplify training. In addition, our modeling interface provides us with calibrated cameras by construction, since we know from which viewpoint each drawing is created. Unrolling the network allows us to apply the camera transformations explicitly as we iterate over each viewpoint. Ji et al.~\shortcite{Ji_2017_ICCV} describe a multi-view reconstruction network that fuses two aligned voxel grids, each being filled with the color rays originating from the pixels of two calibrated input views. Their method extends to more than two views by averaging the predictions given by multiple pairs of views. Our updater network follows a similar strategy of implicitly encoding the camera orientation in the voxel grid. However, we iterate our updater over all drawings, one at a time, rather than combining multiple pairwise predictions at once. This design choice makes our method more sensitive to the order if which the drawings are created.

While CNNs have been mostly applied to photographs, they have also demonstrated impressive performances on tasks similar to ours, such as sketch cleanup \cite{SimoSerraSIGGRAPH2016}, sketch colorization \cite{sangkloy2016scribbler}, sketch-based retrieval \cite{wang2015sketch,Sangkloy2016}, and sketch-based modeling of parametric shapes \cite{gen2016interactive,huang2016shape,HanGY17}. CNNs have also recently achieved promising results in \emph{synthesizing} images \cite{YanYSL16,DB16d,tvsn_cvpr2017,Chen2017} and even 3D models \cite{3dShapeNets15,DTB16,3dgan16,li_sig17,Fan_cvpr2017} from low-dimensional feature vectors and attributes. We pursue this trend by training a deep network to generate voxelized objects from line drawings, offering precise user control on the shape being generated. 

Two recent methods with similar goals have been developed concurrently to ours. First, Liu et al. \shortcite{Liu2017} combine a voxel sculpting interface with a generative network to project the coarse voxel shapes modeled by the user onto a manifold of realistic shapes. We see our sketch-based interface as an alternative to voxel-sculpting. Second, Lun et al. \shortcite{Lun2017} propose a method to reconstruct a 3D object from sketches drawn from one to three orthographic views. We share several ideas with this latter work, such as training with synthetic drawings and predicting 3D shapes from multiple views. On the one hand, Lun et al. achieve finer reconstructions than ours by extracting a 3D surface from multiple depth maps rather than from a voxel grid. On the other hand, they train separate networks to process different combinations of front/side/top views, while our updater network allows us to fuse information from any of the $13$ viewpoints available in our interface. In addition, we integrated our approach in an interactive system to demonstrate the novel workflow it enables.

\begin{figure*}[!t]
		\begin{center}
			\includegraphics[width=\textwidth]{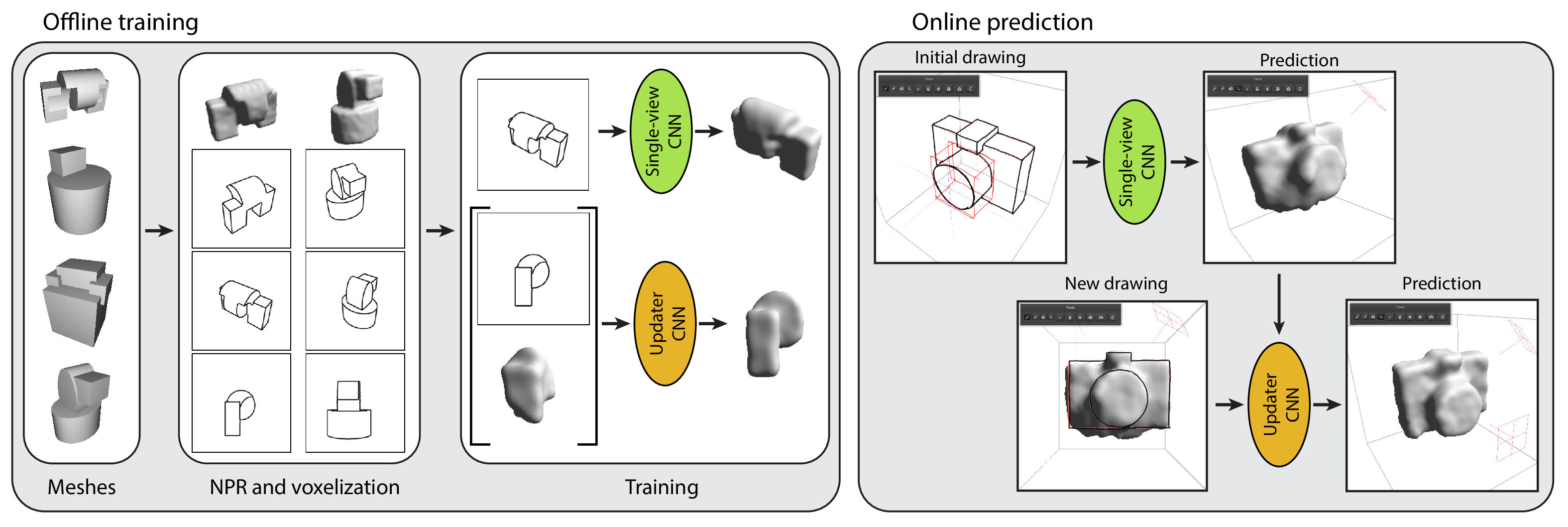}
			\caption{Overview of our method. Left: We train our system with a large collection of 3D models, from which we generate voxel grids and synthetic drawings. We train a single-view CNN to predict an initial reconstruction from a single drawing, as well as an updater CNN to refine a reconstruction given a new drawing. Right: The single-view CNN allows users to obtain a complete 3D shape from a single drawing. Users can refine this initial result by drawing the shape from additional viewpoints. The updater CNN combines all the available drawings to generate the final output.}
			\label{fig:overview}
		\end{center}
	\end{figure*}

\section{Overview}

Figure~\ref{fig:overview} provides an overview of our system and the underlying convolutional neural networks. The left part of the figure illustrates the offline training of the deep neural networks. Given a dataset of 3D models, we first generate a voxel representation of each object, along with a series of line drawings rendered from different viewpoints. Our single-view CNN takes a drawing as input and generates a voxel grid with probabilistic occupancy. Our updater CNN also takes a drawing as input, and complements it with an initial 3D reconstruction provided by the single view network. Note that we transform this reconstruction according to the camera matrix of the input drawing, so that the updater CNN does not have to learn the mapping between the 3D volume and a given viewpoint. The updater network fuses the information from these two inputs to output a new 3D reconstruction. In practice, we repeatedly loop the updater over all available drawings of a shape to converge towards a multi-view coherent solution. 

The right part of Figure~\ref{fig:overview} illustrates our online modeling workflow. The main motivation of our approach is to provide a workflow that seamlessly combines 2D sketching and 3D visualization. At the beginning of a modeling session, our interface displays an empty 3D space seen from a 3/4 view. We additionally display perspective guidance to help users draw with the same perspective as the one used to generate the training data, as detailed in Section~\ref{sec:UI}. Once an initial drawing is completed, the user can invoke our single-view CNN to obtain its 3D reconstruction, which she can visualize from any viewpoint. The user can then refine the shape by re-drawing it from a new viewpoint, using the current reconstruction as a reference. We feed each new drawing to the updater network to generate an improved 3D reconstruction.

	
	

\section{Volumetric prediction from line drawings}
	The key enabler of our modeling system is a deep convolutional network that we train to predict voxelized objects from line drawings. We first present our single-view network that takes as input one drawing to generate an initial 3D reconstruction. We then introduce our \emph{updater network} that iteratively fuses multi-view information by taking as input a drawing and an existing volumetric prediction. We illustrate our network in Figure~\ref{fig:network} and provide a detailed description in supplemental materials. We discuss and compare our design choices against alternative solutions in Section~\ref{sec:eval}. 
	

\begin{figure}[!t]
	\begin{center}
		\includegraphics[width=\linewidth]{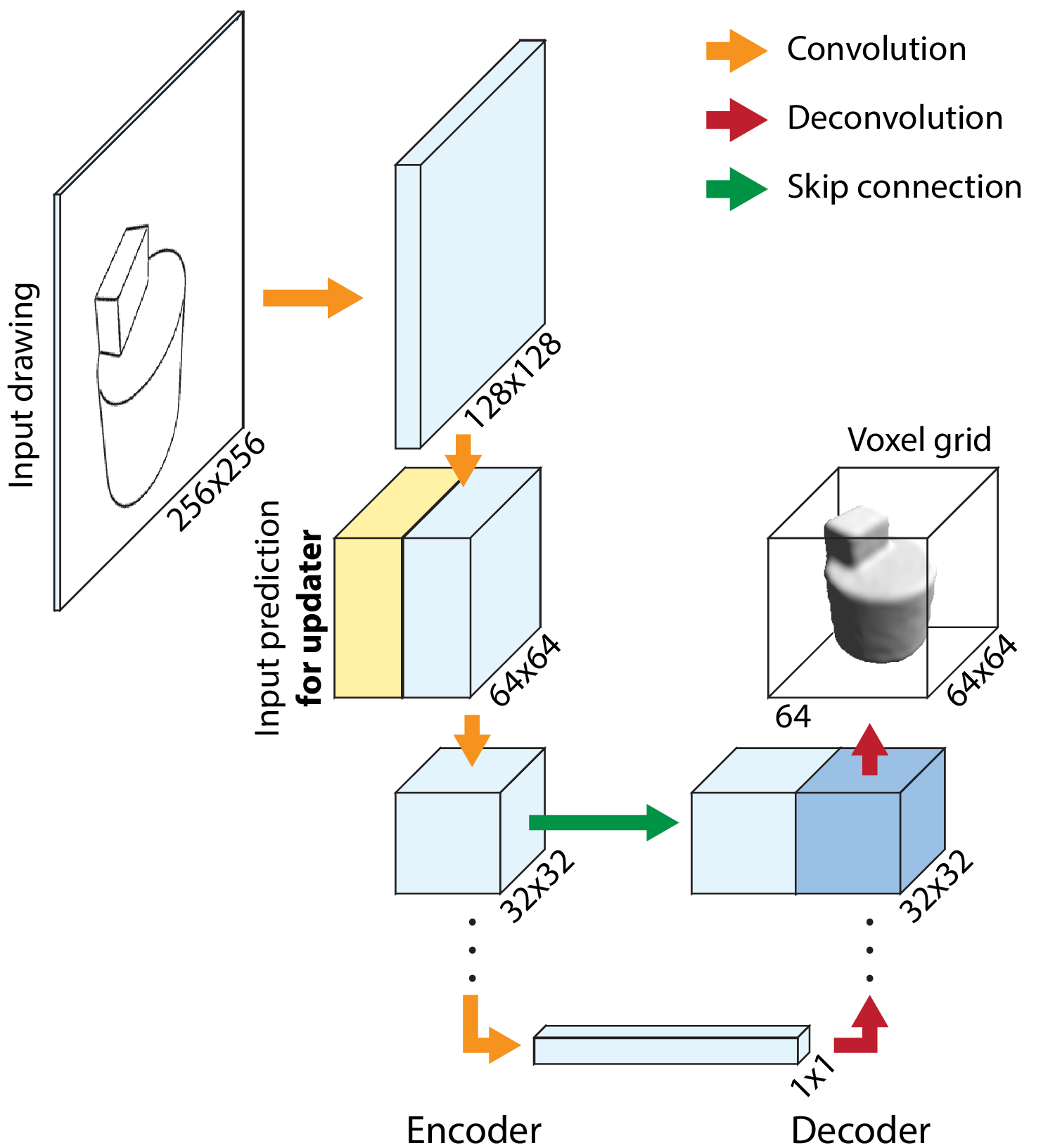}
		\caption{Our network follows a so-called ``U-net'' encoder-decoder architecture. The input drawing is processed by a series of convolution and rectified linear units to extract high-dimensional features at low spatial resolution. These features are then processed by deconvolutions and rectified linear units to generate the multi-channel image that represents our voxel grid. Skip connections, shown in green, concatenate the output of convolutional layers to the output of deconvolutional layers of the same resolution. These connections allow high-resolution features to bypass the encoder bottleneck, allowing the network to exploit multi-scale information for decoding. The updater network also takes an existing prediction as input, shown in yellow.}
		\label{fig:network}
	\end{center}
\end{figure}

	\subsection{Single view prediction} \label{sec:single_view}
	Our single-view network follows an encoder-decoder architecture typical of image generation tasks such as depth prediction \cite{eigen2015predicting}, colorization \cite{sangkloy2016scribbler}, novel view synthesis \cite{tvsn_cvpr2017}. The encoder passes the input image through a series of convolutions of stride $2$ and rectified linear units to progressively reduce spatial resolution while increasing feature dimensionality, effectively extracting a compact representation of the image content. The decoder passes this representation through a series of deconvolutions of stride $2$ and rectified linear units to progressively generate a new visualization of the image content, in our case in the form of a voxel grid. 
	
Following \cite{RonnebergerFB15}, we also include skip connections between the encoder and decoder layers of equal resolution. These skip connections allow local information to bypass the encoder bottleneck, providing the decoder with multi-scale features that capture both global context and fine image details. Isola et al.~\shortcite{pix2pix2016} have demonstrated the effectiveness of a similar ``U-net'' architecture for image-to-image translation tasks.

The task of our network is to classify each voxel as occupied or empty. We model the voxel grid as a multi-channel image, where each channel corresponds to one depth slice. Given this representation, our network can be seen as an extension of existing depth prediction networks \cite{eigen2015predicting}, where we not only predict the depth of the visible surface but also all occluded voxels along the viewing ray corresponding to each pixel of the input drawing. Since our modeling interface employs a perspective camera model, the voxel grid associated to a drawing actually forms a pyramid in 3D space. 
While we considered using an orthographic camera model for simplicity, our early experiments suggest that perspective cues significantly help the network to predict depth for regular shapes.

	\subsection{Multi-view prediction}
	\label{sec:updater}
Our updater network adopts a similar architecture as the one described above, except that it also takes as input an existing volumetric prediction and uses the input drawing to refine it. In practice, we concatenate the existing prediction with the output of the second convolution layer, as illustrated in Figure~\ref{fig:network} (yellow block). 
Note that we do not threshold the probabilities of occupancy in the existing prediction, which allows the updater network to account for the uncertainty of each voxel.

\begin{figure}[!t]
		\begin{center}
		\includegraphics[width=\linewidth]{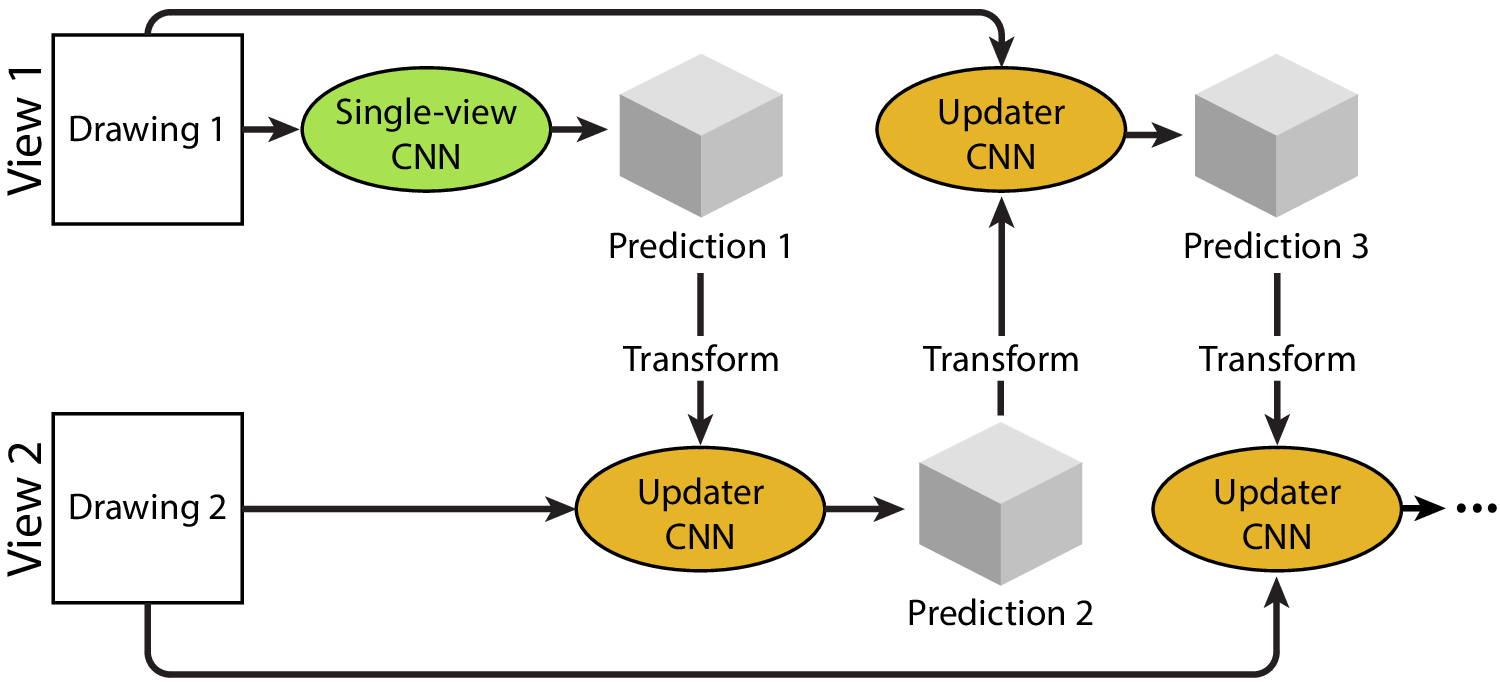}
		\caption{We apply the updater network iteratively, alternating between views to converge towards a multi-view coherent solution. Here we illustrate a few iterations between two views, although we loop over more views when available.}
		\label{fig:schedule_update}
		\end{center}
	\end{figure}

\paragraph{Iterative update.}
The updater network processes one drawing at a time, which allows us to handle an unbounded number of views. However, each update may modify the prediction in a way that is not coherent with the other views. We found that we can achieve multi-view consistency by iteratively applying the updater network until convergence, akin to a coordinate descent optimization. Figure~\ref{fig:schedule_update} illustrates this process with two views: the first drawing is given as input to the single-view network to generate a first prediction. This prediction is then given to the updater network along with the second drawing to produce a refined solution. The resulting voxel grid can now be processed again by the updater, this time taking the first drawing as input. This process generalizes to more views by looping the updater over all drawings in sequence. 
In practice, we used $5$ iterations for all results in the paper. We evaluate the convergence of this iterative scheme in Section~\ref{sec:eval}.

\begin{figure}[!t]
	\begin{center}
		\includegraphics[width=\linewidth]{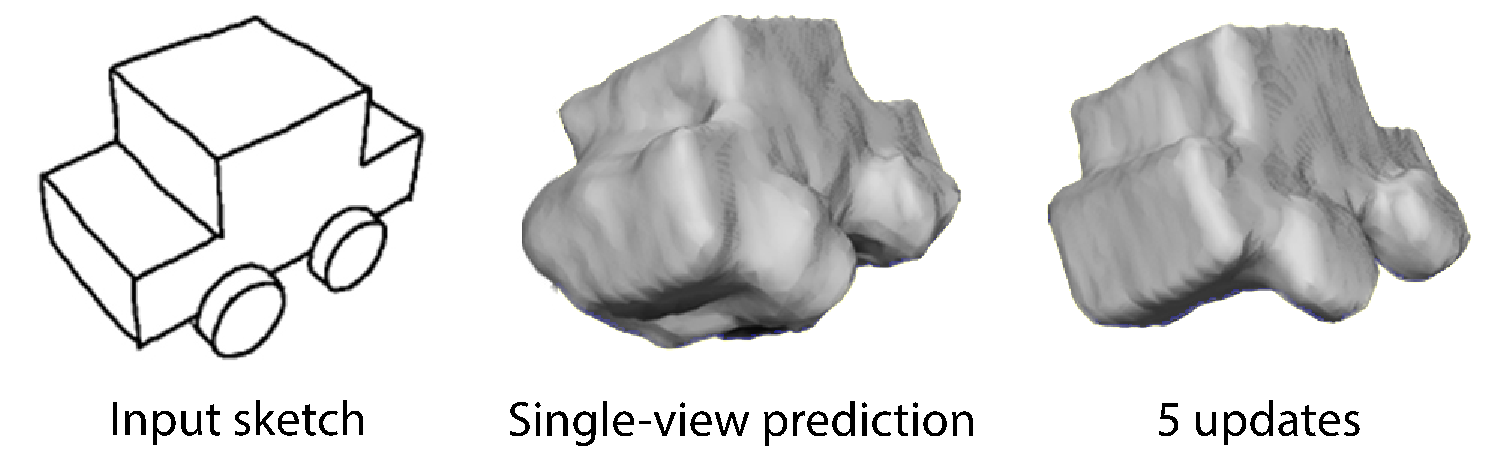}
		\caption{The updater network can refine the prediction even when only one drawing is available.}
		\label{fig:single_view_refine}
	\end{center}
\end{figure}

\paragraph{Resampling the voxel grid.}
As mentioned in Section~\ref{sec:single_view}, we designed our networks to process and generate voxel grids that are expressed in the coordinate system of the input drawing. When dealing with multiple drawings, the prediction obtained with any drawing needs to be transformed and resampled to be passed through the updater network with another drawing. In practice, we store the prediction in a reference voxel grid in world coordinates, and transform this grid to and from the coordinate system of each drawing on which we run the updater network.

\paragraph{Single-view refinement.}
While we designed the updater network to fuse information between multiple views, we found that it is also able to refine a single-view prediction when used as a feedback loop on a single drawing, as shown in Figure~\ref{fig:single_view_refine}. This observation may seem counter-intuitive, since the updater does not have more information than the single-view network in that configuration. We hypothesize that iterating the updater on a single drawing emulates a deeper network with higher capacity. Note also that a similar iterative refinement has been demonstrated in the context of pose estimation \cite{carreira2015human,oberweger2015training}.

	
	 
 

\section{Data generation and training}

The two CNNs outlined above require pairs of drawings and ground truth voxel grids for training. Ideally, the training data should be representative of the distribution of drawings and 3D shapes that users of our system are likely to create. However, while datasets of cartoon drawings of objects have been collected via crowd-sourcing \cite{eitz2012sbsr}, building a dataset of perspective drawings registered with 3D models raises many additional challenges. In particular, we assume that users of our system are sufficiently skilled to draw 3D objects in approximate perspective, which may not be the case for the average Mechanical Turk worker~\cite{eitz2012hdhso}. In addition, crowd-sourcing such a drawing task would require significant time and money, which prevents iterative design of the dataset, for instance to adjust the complexity and diversity of the 3D shapes.

Similarly to recent work on sketch-based procedural modeling \cite{gen2016interactive,huang2016shape}, we bypass the challenges of real-world data collection by generating our training data using non-photorealistic rendering. This approach gives us full control over the variability of the dataset in terms of shapes, rendering styles, and viewpoints.

\begin{figure}[!t]
	\begin{center}
		\includegraphics[width=\linewidth]{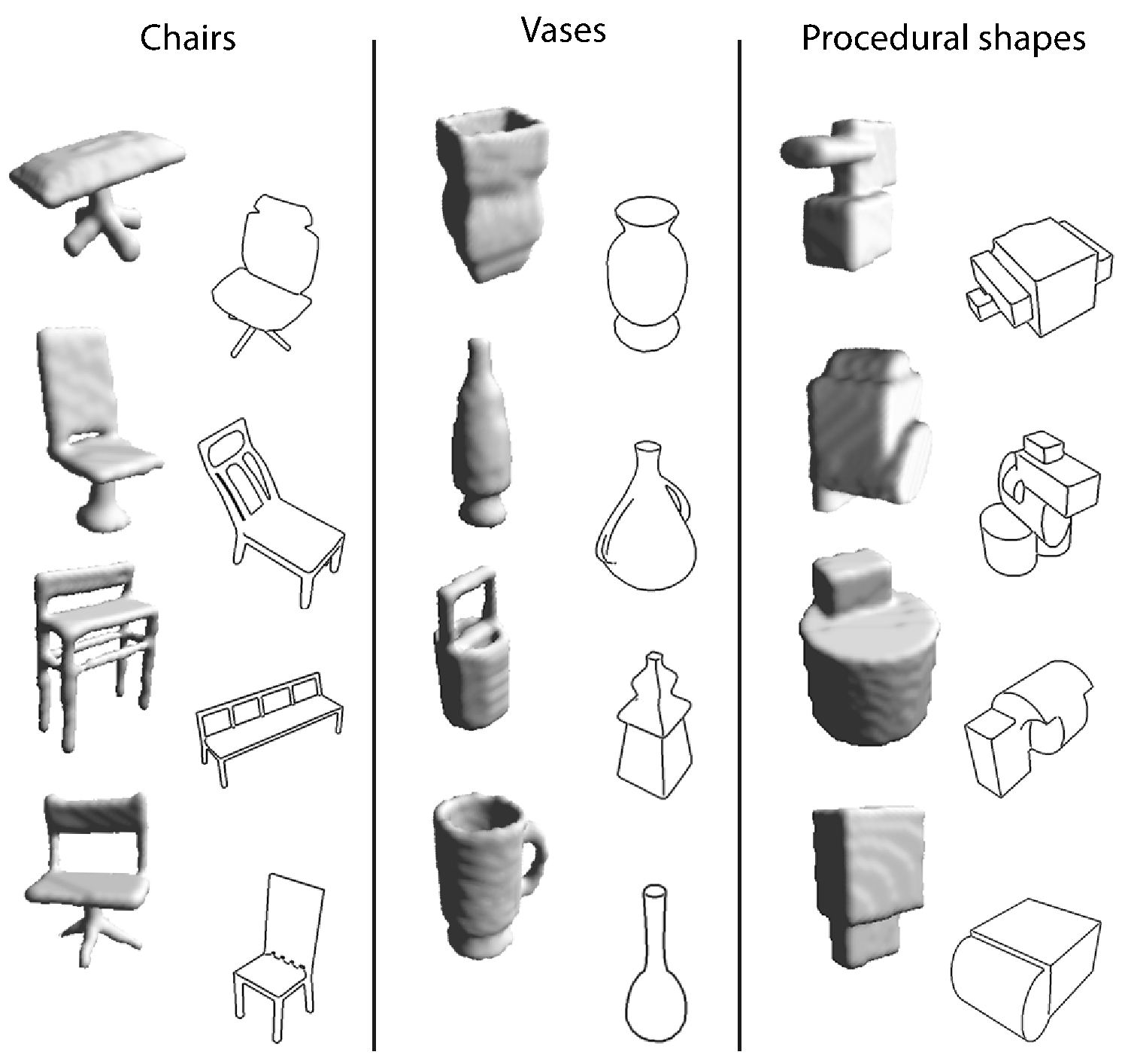}
		\caption{Representative voxelized objects and drawings from our three datasets.}
		\label{fig:datasets}
	\end{center}
\end{figure}

\subsection{3D objects}
The main strength of our data-driven approach is its ability to capture the characteristic features of a class of objects. We experimented with two sources of 3D object datasets: online repositories and shape grammars.

\paragraph{Online repositories.}
A first usage scenario of our system is to train it with specific object classes. For instance, a furniture designer could train the system with chairs and tables, while a car designer could train the system with various vehicles. In practice, we tested this approach with the two largest classes of the ShapeCOSEG dataset \cite{Active2012}, which contain $400$ chairs and $300$ vases, some of which are shown in Figure~\ref{fig:datasets}. For each dataset, we used $90\%$ of the objects for training and the other $10\%$ for testing.

\paragraph{Shape grammars.}
One drawback of online shape repositories is that they are dominated by a few object classes, such as tables, chairs, cars and airplanes \cite{shapenet2015}, which may not cover the types of objects the user wants to model. In addition, many of these objects are very detailed, while we would like our system to also handle simple shapes to allow coarse-to-fine explorative design. We address these limitations by training our system with abstract shapes with varying degrees of complexity.

We designed a simple procedure that generates shapes by combining cubes and cylinders with CSG operations. Our procedure iteratively constructs a shape by adding or substracting random primitives. At each iteration, we position the new primitive on a random face of the existing shape, scale it by a random factor in each dimension and displace it by a small random vector while maintaining contact. The primitive is either merged with or subtracted from the existing shape. We inject high-level priors in this procedure by aligning each primitive with one of the three world axes, and by symmetrizing the shape with respect to the xy plane in world coordinates. The resulting axis-aligned, symmetric shapes resemble man-made objects dominated by flat orthogonal faces, yet also contain holes, concavities and curved parts. We generated $20,000$ random shapes with this procedure, some of which are shown in Figure~\ref{fig:datasets}. We isolated $50$ of these shapes for testing, and used the rest for training.

 
\paragraph{Voxelization.}
We voxelize each object at a resolution of $64^3$ voxels using Binvox \cite{Nooruddin03,binvox16}. We scale each object so that the voxel grid covers $120\%$ of the largest side of the object's bounding box.

\subsection{Line rendering}
We adopt the image-space contour rendering approach of Saito and Takahashi~\shortcite{Saito1990}, who apply an edge detector over the normal and depth maps of the object rendered from a given viewpoint. Edges in the depth map correspond to depth discontinuities, while edges in the normal map correspond to sharp ridges and valleys. We render each drawing at a resolution of $256^2$ pixels.

An exciting direction of future work would be to train our system with other rendering techniques such as suggestive contours~\cite{DeCarlo2003} and hatching~\cite{hertzmann2000illustrating} to cover a wider range of styles.


\subsection{Viewpoints}
Viewpoint is a major source of ambiguity for line drawing interpretation. We now describe our strategies to significantly reduce ambiguity for the single-view network by restricting camera orientation and position. We relax these restrictions for the updater network since it can handle more ambiguity thanks to the existing prediction it also takes as input.

\paragraph{Camera orientation.}
Representing a 3D object with a single drawing necessarily induces ambiguity. The design literature \cite{Eissen11} as well as other sketching systems \cite{bae2008ilovesketch,shao2012crossshade,Xu2014} recommend the use of ``informative''  
\setlength{\columnsep}{0.1in}
\begin{wrapfigure}[6]{r}{0.3\linewidth}
	\vspace{-0.2in}
	\includegraphics[width=\linewidth]{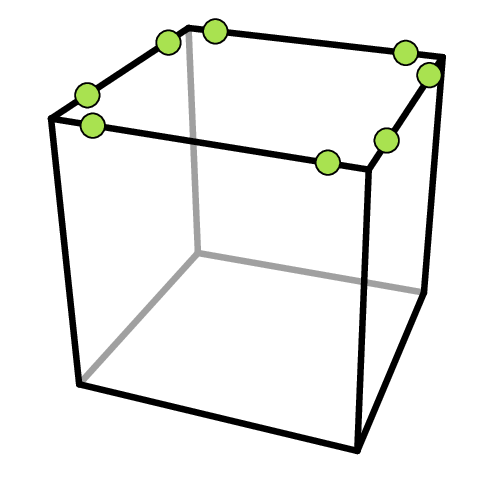}
	\vspace{-0.3in}
\end{wrapfigure}
perspective viewpoints 
that reduce ambiguity by showing the 3D 
object with minimal foreshortening on all sides. 
We follow this practice to train our single-view network. We render each object from eight viewpoints positioned near the top corners of its bounding box, as shown in inset.

In addition, designers frequently adopt so-called ``accidental'' viewpoints when representing a shape with several drawings, such as the common front, side and top views. We include these viewpoints in the training set of our updater network since we found   
them useful to refine axis-aligned shapes. However, we do not use 
\setlength{\columnsep}{0.1in}
\begin{wrapfigure}[6]{r}{0.3\linewidth}
\vspace{-0.15in}
\includegraphics[width=\linewidth]{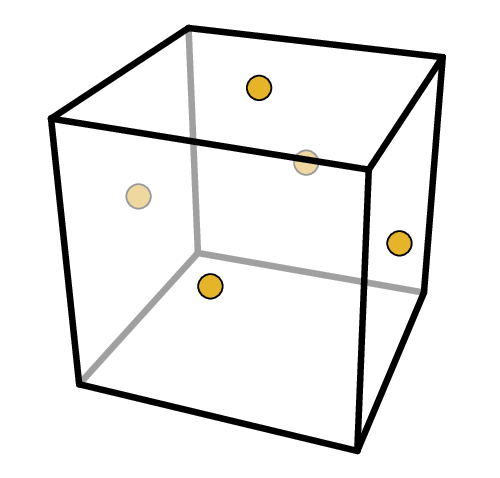}
\vspace{-0.3in}
\end{wrapfigure}
these viewpoints with the single-view network because they often yield significant occlusions, which make them very challenging to interpret in the absence of additional information. The inset shows the additional viewpoints available to the updater network.

\paragraph{Camera position.}
Line drawings also have an inherent depth ambiguity: the same drawing can represent a small object close to the camera, or a big object far from the camera. We reduce such ambiguity for the single-view network by positioning the 3D object at a constant distance to the camera. In addition, we achieve invariance to 2D translations in the image plane by displacing the camera by a random vector perpendicular to the view direction. 

However, a 2D translation in one view potentially corresponds to a translation in depth in another view, which prevents us imposing a constant distance to the camera for the updater network. We thus train the updater network with random 3D displacements of the camera. We found that the updater network succeeds in exploiting the existing prediction to position the object in depth.

\subsection{Training procedure}	
\begin{figure}[!t]
	\begin{center}
		\includegraphics[width=\linewidth]{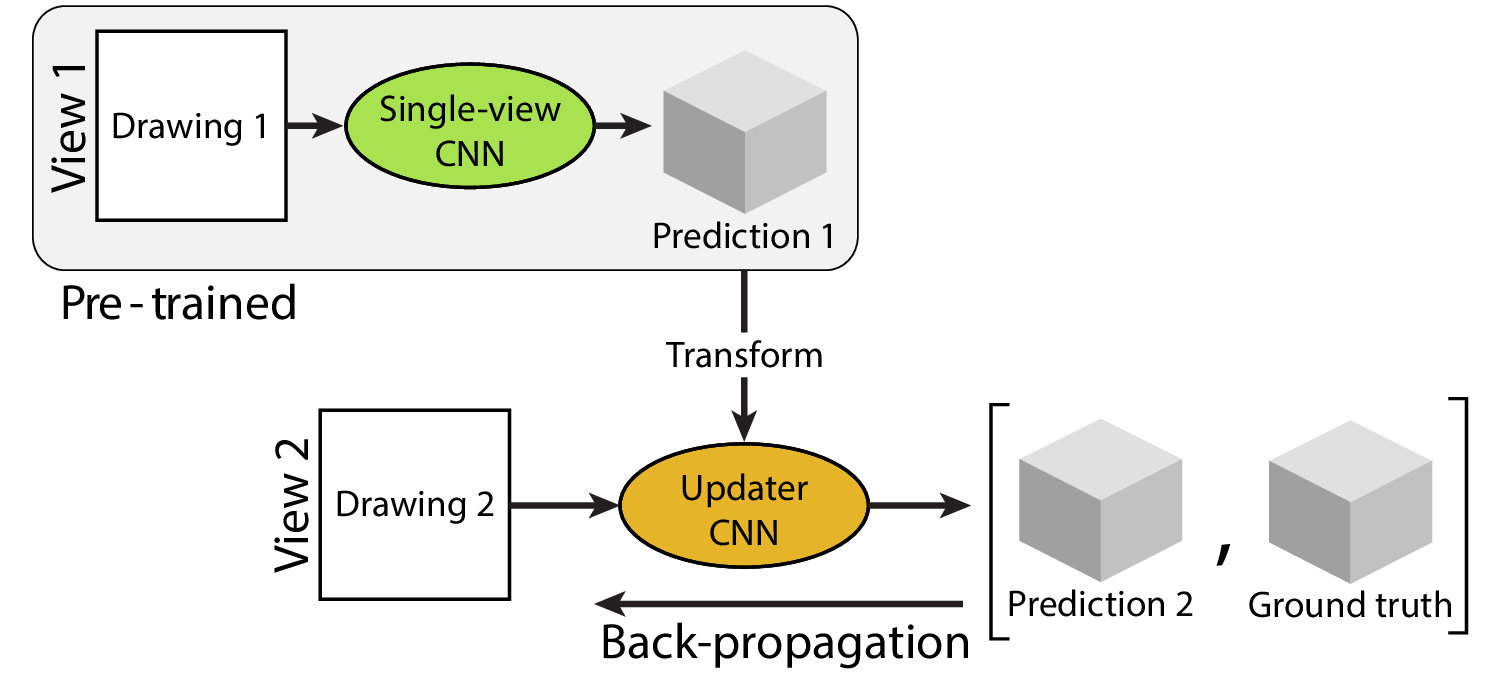}
		\caption{We first train our single-view network on ground truth data, then use its predictions as training data for the updater network.}
		\label{fig:updater}
	\end{center}
\end{figure}

We train our single view network by providing a line drawing as input and a ground truth voxel grid as output. However, training our updater network is more involved since we also need to provide an existing prediction as input. Given a drawing and its associated 3D model, we obtain an initial prediction by running the single-view network on another viewpoint of the same object. Figure~\ref{fig:updater} illustrates this process. We thus need to train the single-view network before training the updater.

We trained our system using the Adam solver \cite{KingmaB14}, using batch normalization \cite{ioffe2015batch} to accelerate training. We fixed Adam's parameters to $\beta_1 = 0.5$, $\beta_2 = 0.999$, $\epsilon = 1e-8$. We fixed the learning rate to $0.0002$ and trained the networks for $1,000,000$ iterations. Training the complete system took around a week on a NVidia TitanX GPU.

\section{User interface}
\label{sec:UI}
	
\begin{figure}[!t]
		\begin{center}
			\includegraphics[width=\columnwidth]{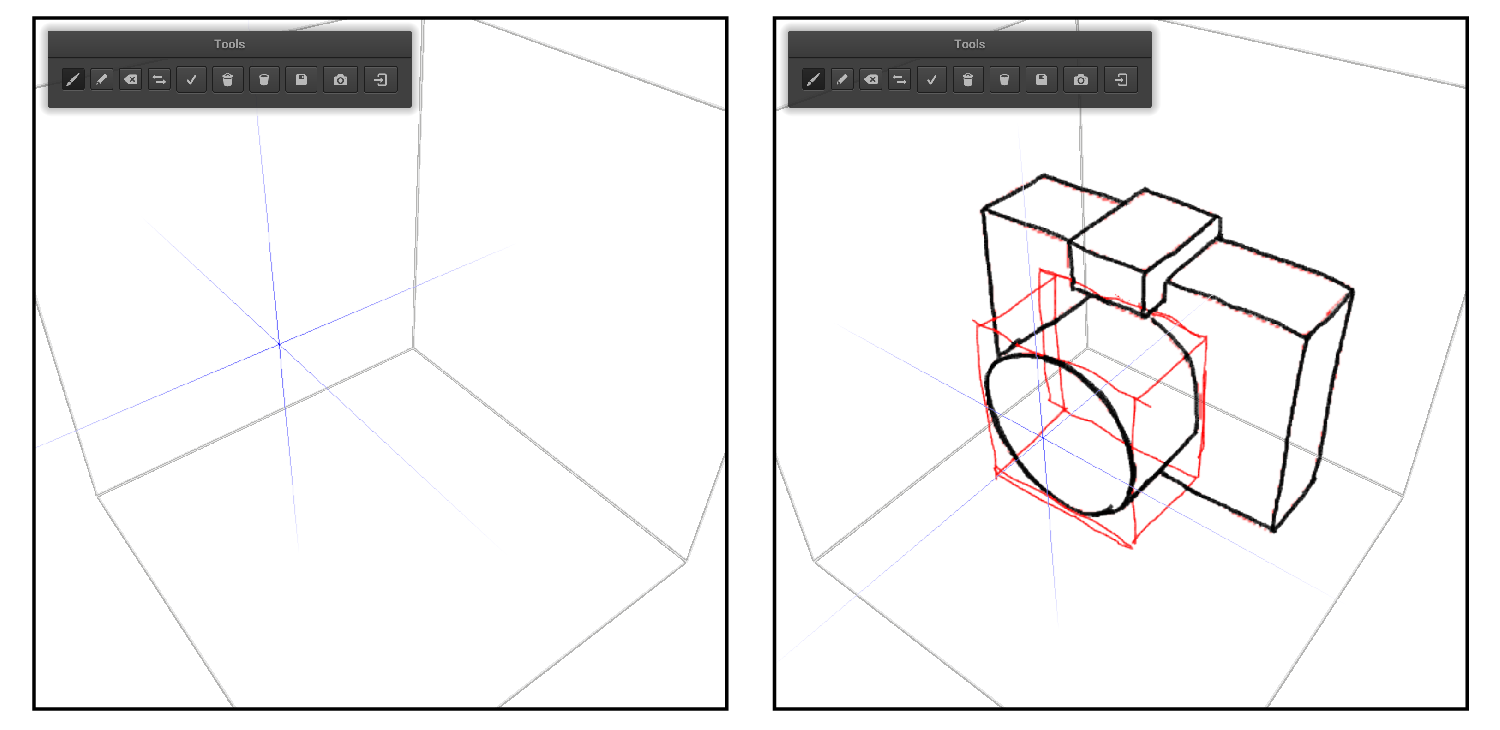}
			\vspace{-6mm}
			\caption{Screenshots of our user interface. We display axis-aligned lines around the cursor to guide perspective drawing (left, shown in blue). We also allow users to draw construction lines (right, shown in red). Only the black lines are processed by our 3D reconstruction engine.}
			\vspace{-2mm}
			\label{fig:interface}
		\end{center}
	\end{figure}

Figure~\ref{fig:interface} shows the interactive interface that we built around our deep 3D reconstruction engine. We designed this interface to reproduce traditional pen-on-paper freehand drawing. However, we introduced several key features to guide users in producing drawings that match the characteristics of our training data in terms of viewpoints and perspective.

	Similarly to the seminal Teddy system \cite{igarashi1999teddy}, the working space serves both as a canvas to draw a shape and as a 3D viewer to visualize the reconstruction from different viewpoints. While we allow free viewpoint rotations for visualization, we restrict the drawing viewpoints to the ones used for training. In particular, we impose a 3/4 view for the first drawing, and snap the camera to one of the $13$ viewpoints available to the updater for subsequent drawings.
	
	The menu in the top left allows users to switch from 2D drawing to 3D navigation and also provides basic drawing tools (pen and eraser). In addition, we provide a ``construction line'' mode to draw scaffolds \cite{schmidt2009analytic} and other guidance that will not be sent to the network (shown in red in our interface). We found such lines especially useful to lay down the main structure of the object before drawing precise contours (shown in black). We further facilitate perspective drawing by displaying three orthogonal vanishing lines centered on the pen cursor (shown in blue) and by delineating the working space with a wireframe cube.
	
	
	For each voxel, our networks estimate the probability that it is occupied. We render the shape by ray-casting the $0.5$ iso-surface of this volume, using the volumetric gradient to compute normals for shading. We also export the shape as a triangle mesh, which we obtain by apply a marching cube \cite{Lorensen1987} followed by a bilateral filter to remove aliasing \cite{TFM2003}.

\begin{figure*}[!t]
	\begin{center}
		\includegraphics[width=\textwidth]{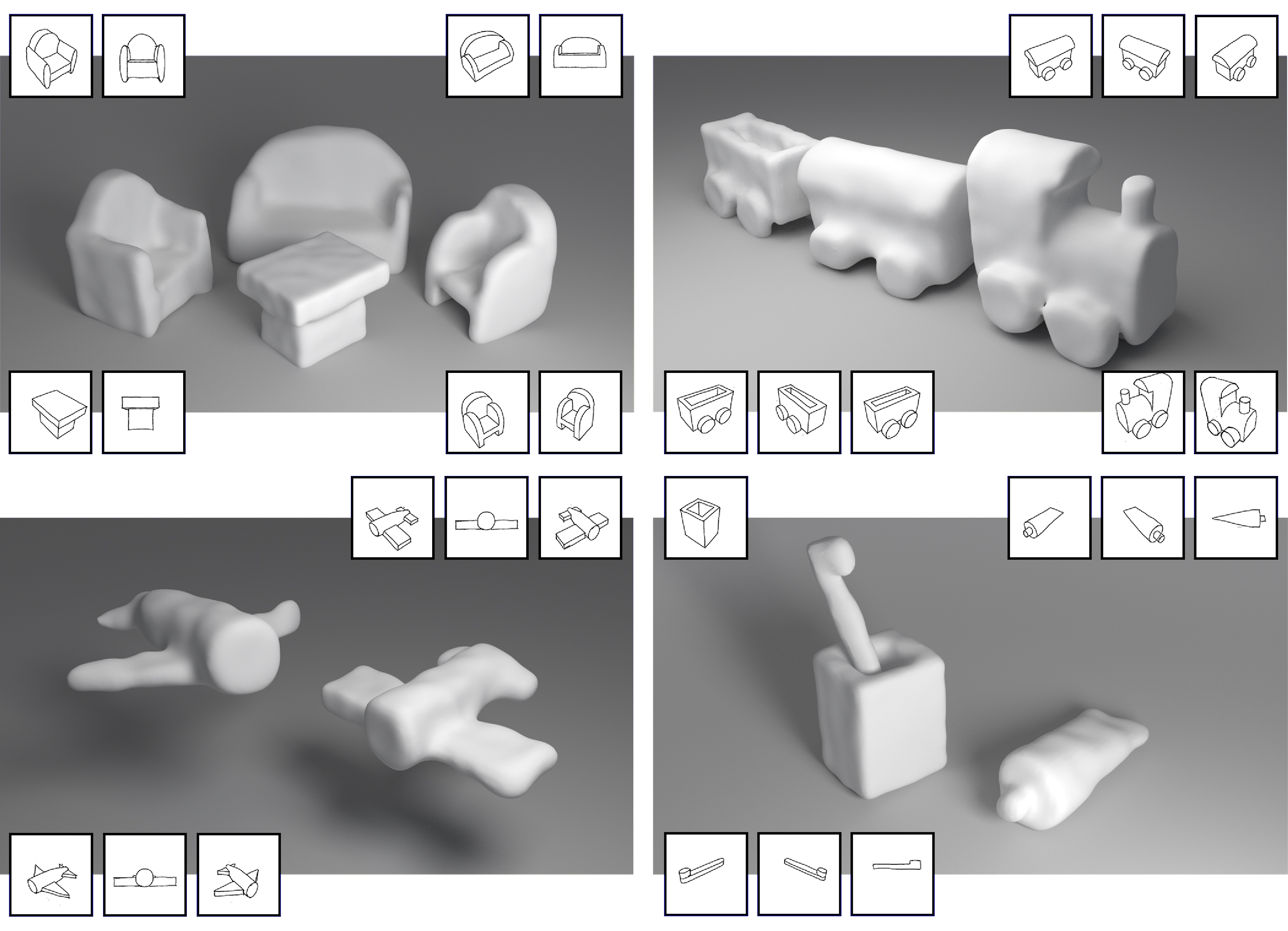}
		\caption{3D scenes modeled using our system. Each object was modeled with two to three hand drawings, shown in insets.}
		\label{fig:rendering}
	\end{center}
\end{figure*}

\begin{figure}[!t]
	\begin{center}
		\includegraphics[width=\linewidth]{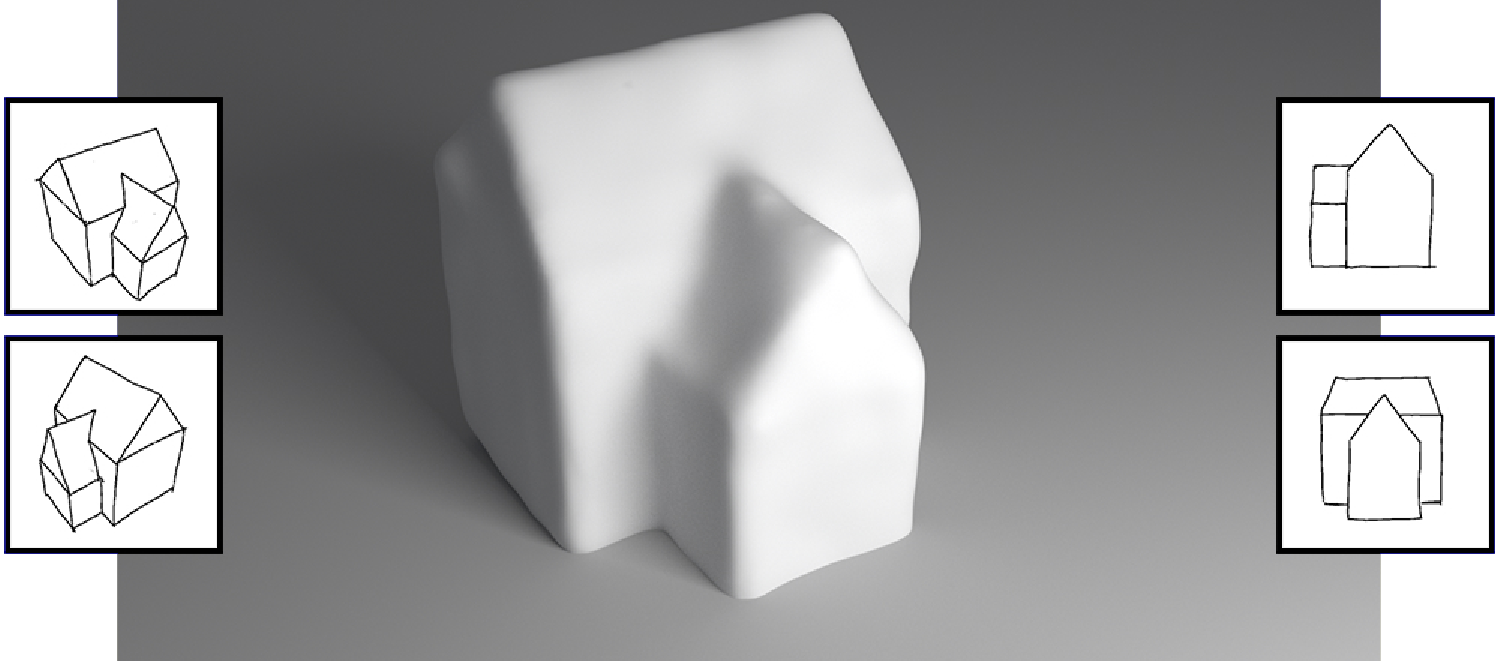}
		\caption{Our system manages to reconstruct the slanted roof of this house, even though it was only trained on shapes composed from axis-aligned cuboids and cylinders.}
		\label{fig:house}
	\end{center}
\end{figure}

\section{Results and evaluation}
\label{sec:eval}

We now evaluate the expressivity and robustness of our method and compare it to alternative approaches. We use the dataset of abstract procedural shapes for these comparisons, and separately evaluate the impact of the other datasets on our method. All results were obtained with a voxel grid of resolution $64^3$.

In all cases we evaluate the quality of a volumetric reconstruction against ground-truth using the intersection-over-union (IoU) metric, which computes the ratio between the intersection and the union of the two shapes \cite{hspHane17,RieglerUBG17}. The main advantage of this metric over the classification accuracy is that it ignores the many correctly-classified empty voxels far away from the shapes. 

\begin{figure*}[!t]
	\begin{center}
		\includegraphics[width=\textwidth]{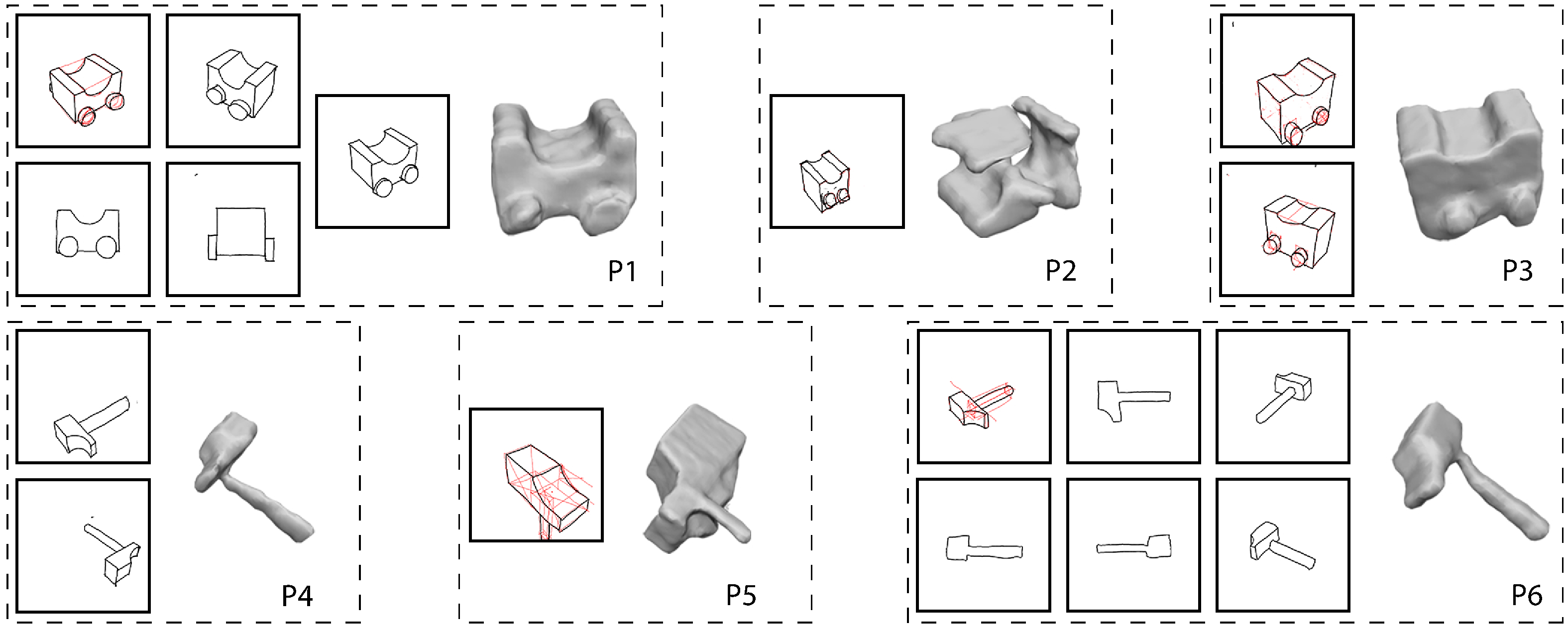}
		\caption{Drawings and 3D objects created by our six novice participants. P1 and P6 obtained the best results by drawing the object in the center of the canvas, with proper perspective. In contrast, P2 drew the object too small and with too approximate perspective to be reconstructed by our system, while P5 left too little room for the handle of the hammer.}
		\label{fig:user_study_res}
	\end{center}
\end{figure*}

\subsection{Creative modeling by experts}
Figure~\ref{fig:rendering} presents several 3D scenes modeled with our system by two expert users. These results were created with the version trained on abstract procedural shapes, which succeeds in interpreting these drawings of diverse man-made shapes. In particular, the CNNs manage to segment the foreground object from its background, combine information from different drawings to reconstruct occluded parts, create holes and concavities such as on the armchairs and on the last wagon of the train. Figure~\ref{fig:house} shows the more challenging case of a house with a slanted roof, which is well reconstructed even though the networks were only trained with shapes made of axis-aligned cuboids and cylinders.

We provide screen captures of a few modeling sessions in the accompanying video, showing how users iterate between 2D sketching and 3D navigation within a single workspace. In particular, users can draw a complete shape from one viewpoint before rotating the 3D model to continue working on it from another viewpoint. This workflow contrasts with the one of existing sketching systems that require users to decompose the object in simple parts \cite{gen2016interactive} or to provide multiple drawings of the shape before obtaining its reconstruction \cite{Rivers2010}. The accompanying video also shows 3D visualizations of the objects, and the supplementary materials contain the corresponding 3D mesh files.

\subsection{Evaluation by novice users}
\setlength{\columnsep}{0.1in}
\begin{wrapfigure}[11]{r}{0.3\linewidth}
	\vspace{-0.15in}
	\includegraphics[width=\linewidth]{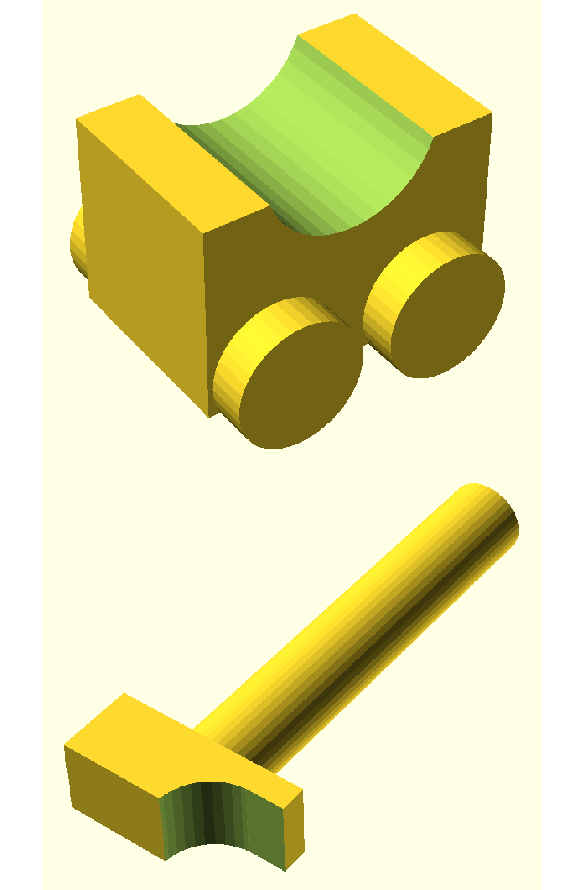}
	\vspace{-0.3in}
\end{wrapfigure}
While we designed our system for artists who know how to draw in perspective, we also conducted a small study to evaluate whether our interface is usable by novices. We recruited six participants with limited drawing and 3D modeling skills (average score of $2.8$ and $2.3$ respectively on a 5-point Likert scale from 1 = \emph{poor} to 5 = \emph{good}). All participants followed a $15$ minutes tutorial to learn how to draw a cube and a cylinder within our interface. We then asked each participant to model one of the two objects shown in inset, which we designed to be expressible by our shape grammar. Figure~\ref{fig:user_study_res} shows the drawings and 3D models they created. 

Overall, participants quickly managed to use our system (average score of $5.5$ on a 7-point Likert scale from 1 = \emph{hard to use} to 7 = \emph{easy to use}). However, many participants were disappointed by the lack of details of the reconstruction and gave an average score of $3.8$ when asked if the 3D model corresponds well to their drawings (1 = \emph{not at all}, 7 = \emph{very well}).    
The best results were obtained by participants who planned their drawings ahead to best represent the shape centered on screen (P1 and P6). In contrast, two participants did not obtain a complete shape because they drew the object too small to capture details (P2) or too big to fit in the drawing area (P5). This observation suggests the need for additional guidance to help novices compose a well-proportioned perspective drawing.

All participants judged the on-cursor vanishing lines helpful to draw in perspective ($6.6$ on average on a 7-point Likert scale from 1 = \emph{not helpful} to 7 = \emph{very helpful}). P3 commented \emph{``Sometimes it seems to me that the guides point to wrong directions, but that is just my sense of perspective that is wrong!''}. All the participants followed the vanishing lines to draw cuboid shapes. However, several participants commented that they would have liked guidance to draw 3D cylinders. In particular, P2 drew very approximate cylinders to represent the wheels of his car, which our system failed to interpret properly.

Finally, even though P1 and P6 created many drawings, several are redundant and did not help our system improve its prediction. We believe that users would interact more effectively with our system if we could indicate which regions of the shape is under-constrained. Recent work on uncertainty quantification in deep networks form a promising direction to tackle this challenge \cite{KendallGal2017}.


\begin{figure}[!t]
	\begin{center}
		\includegraphics[width=\linewidth]{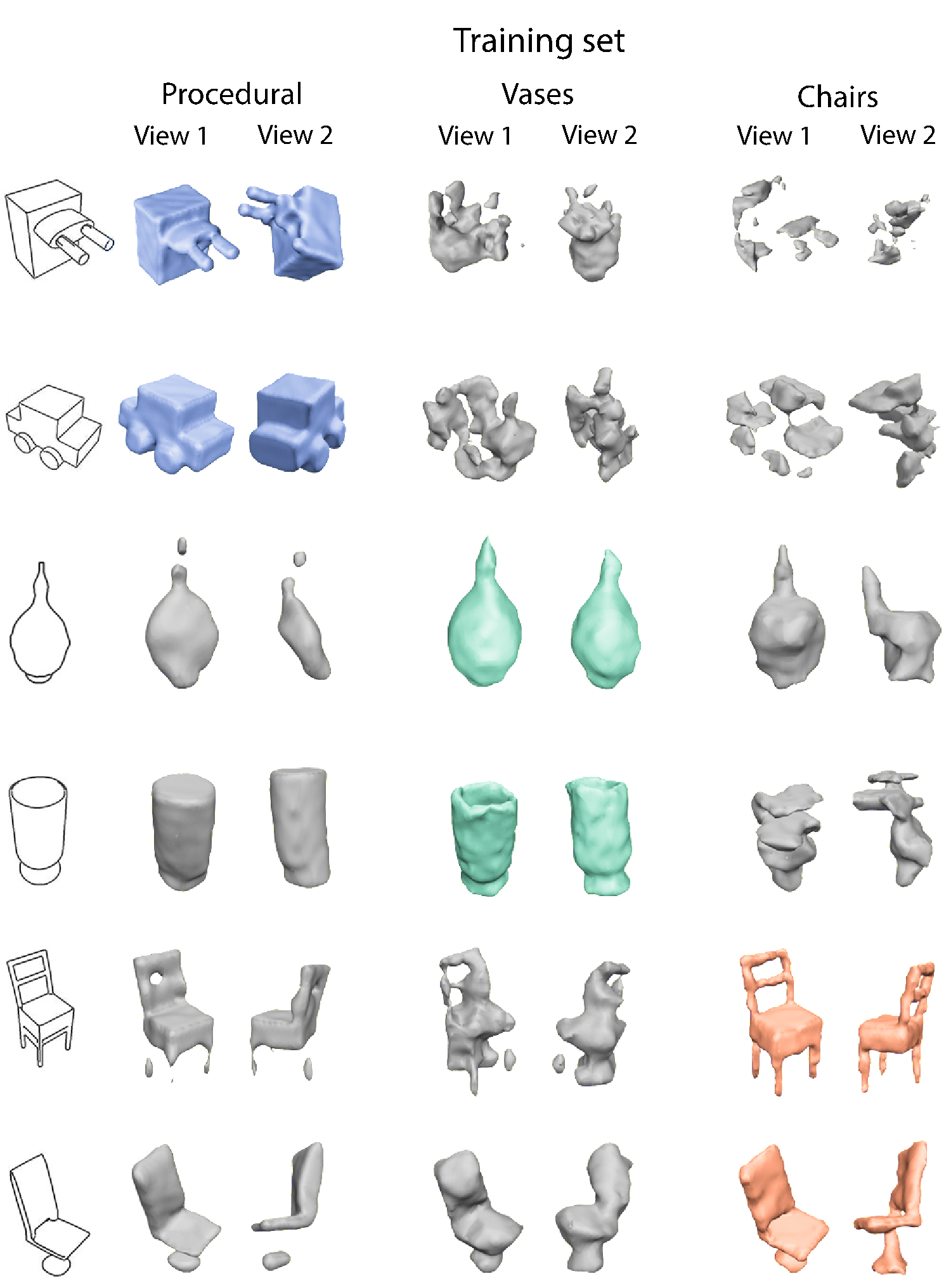}
		\caption{We trained the single-view network with three different datasets, and evaluated each version on the three testing sets. Each version performs best on the testing set for which it was trained, showing that the network learns to specialize to specific object categories. The network performs better on abstract procedural shapes than on chairs and vases, which contain more thin structures and for which the training set is smaller.}
		\label{fig:qual_diff_datasets}
	\end{center}
\end{figure}	

\subsection{Datasets}
\begin{figure}[!t]
	\begin{center}
		\includegraphics[width=0.7\linewidth]{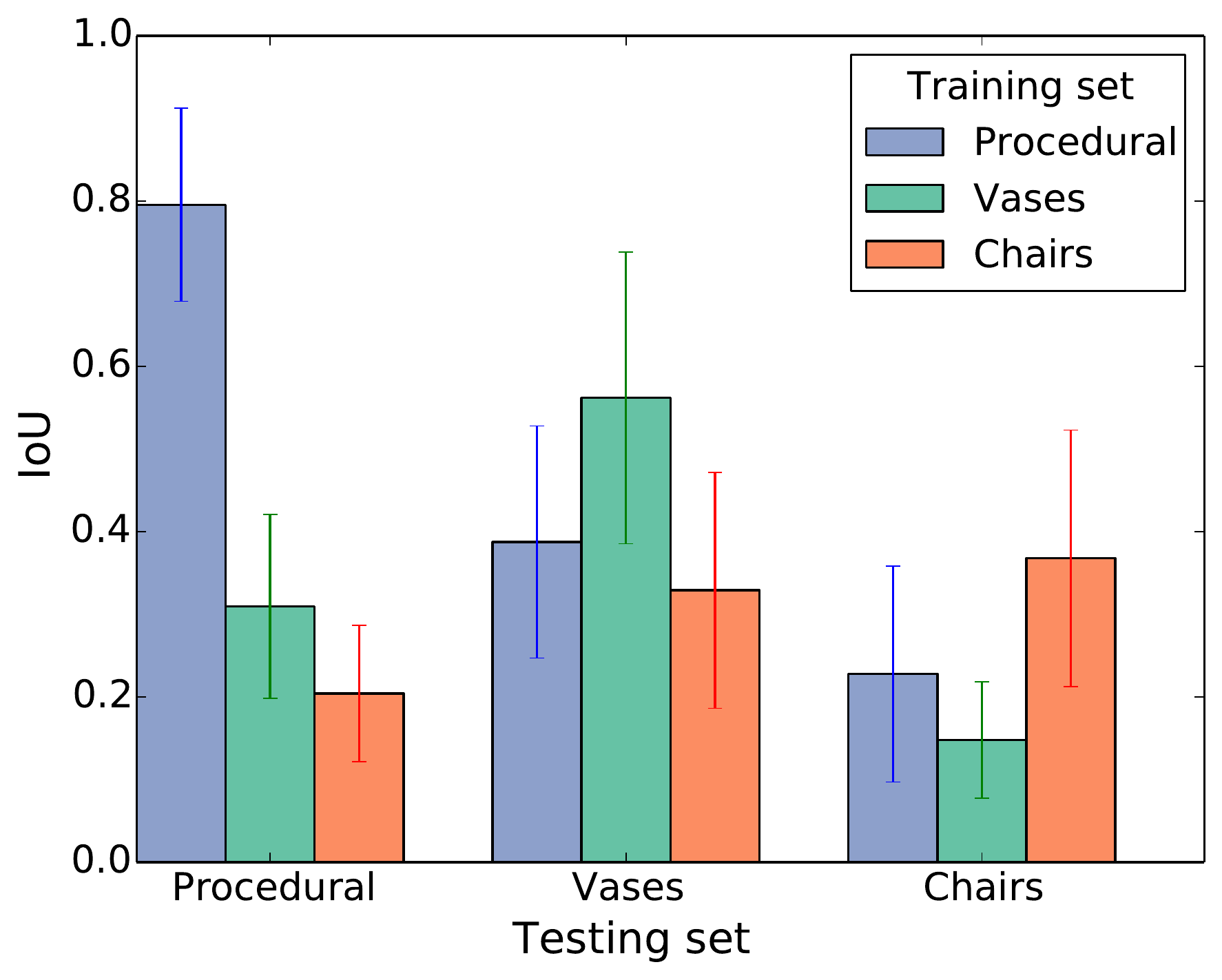}
		\caption{We trained the single-view network with three different datasets (depicted with different colors), and evaluated each version on drawings from the three testing sets (distributed on the x-axis). The network trained on abstract procedural shapes captures the overall shape of objects from other categories, while the networks trained on chairs and vases generalize poorly. Each network performs best on the shapes for which it has been trained.}
		\label{fig:acc_diff_datasets}
	\end{center}
\end{figure}

One of the motivations for our deep-learning-based approach is to allow adaptation to different classes of objects. Figure~\ref{fig:acc_diff_datasets} provides a quantitative evaluation of this ability for the single-view network. This figure plots reconstruction quality over the three testing datasets, using all three training datasets. As expected, the network trained on a given dataset performs best on this dataset, showing its specialization to a specific class. For instance, only the network trained on vases succeeds to create hollow shape from an ambiguous drawing (fourth row, third and fourth column). Interestingly, the network trained on abstract procedural shapes is second best on the other datasets, suggesting a higher potential for generalization. Figure~\ref{fig:qual_diff_datasets} shows representative results for each condition. While the networks trained on chairs and vases manage to reconstruct objects from these classes, they fail to generalize to other shapes. In contrast, the network trained on abstract shapes captures the overall shape of chairs and vases, although it misses some of the details. This superiority of the procedural dataset may be due to its larger size and variability.

\subsection{Convergence of the updater}
\begin{figure}[!t]
		\begin{center}
			\includegraphics[height=0.48\linewidth]{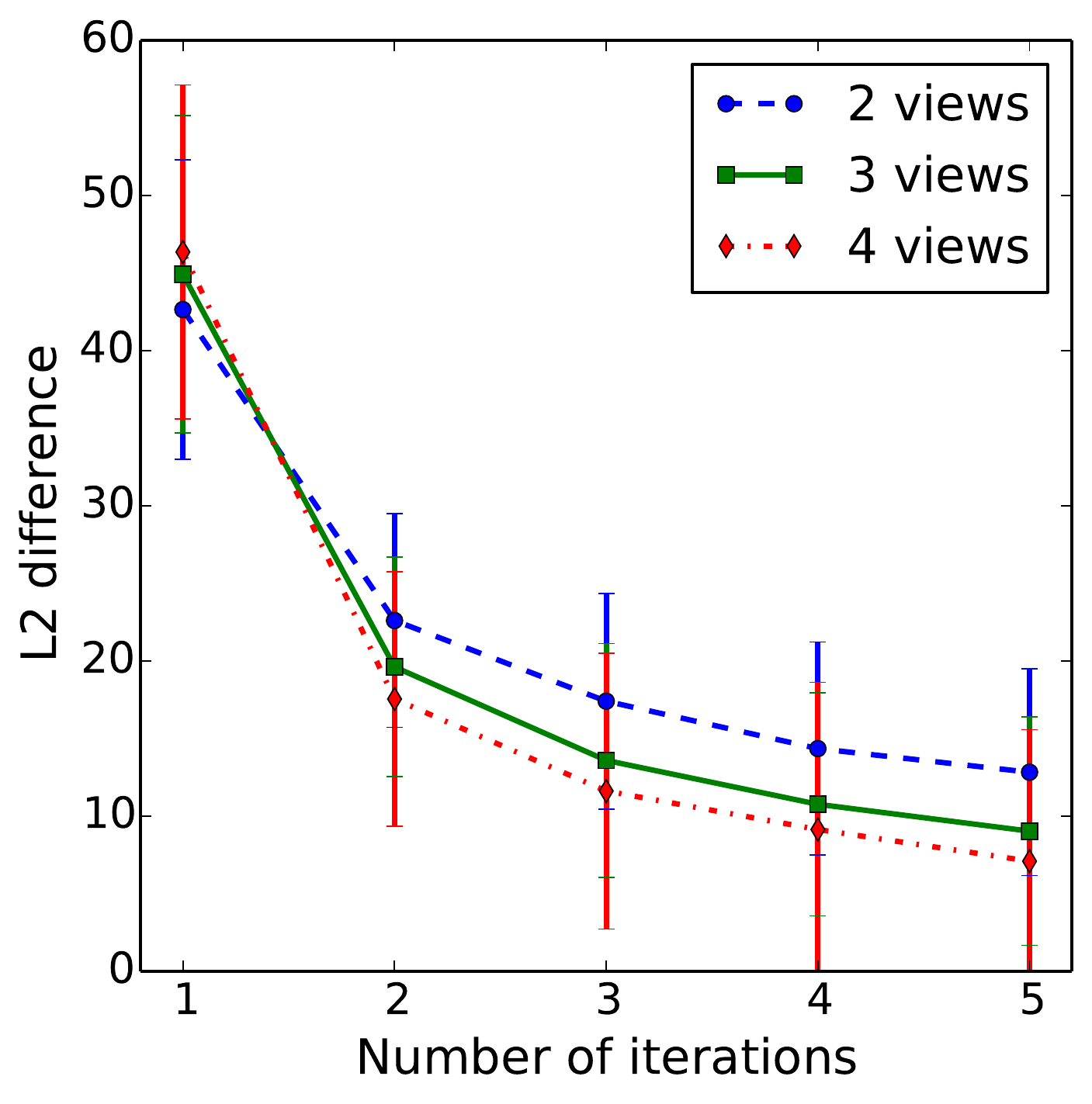}
			\includegraphics[height=0.48\linewidth]{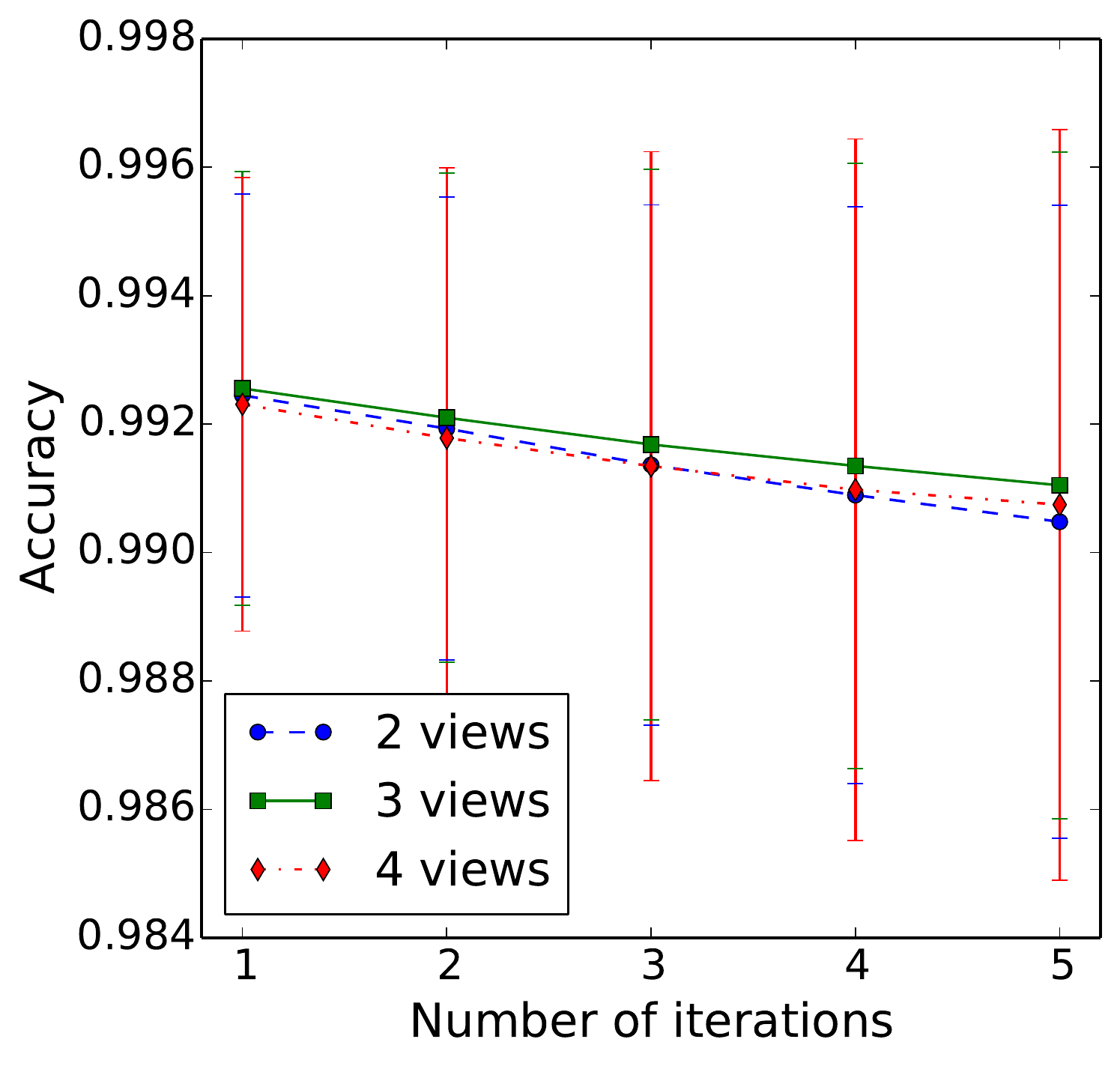}
			\caption{Left: Difference of prediction between successive iterations of the updater network, showing that the network quickly converges towards a stable solution. Right: The accuracy decreases slightly during the iterations. $5$ iterations offer a good trade-off between multi-view coherence and accuracy.}
			\label{fig:plots_convergence}
		\end{center}
	\end{figure}

In what follows, we count one iteration each time the updater network visits all views in sequence. Figure~\ref{fig:plots_convergence}(left) plots the $L^2$ distance between successive iterations, averaged over $50$ abstract shapes rendered from two, three and four random views. While we have no formal proof of convergence, this experiment shows that the algorithm quickly stabilizes to a unique solution. 
However, Figure~\ref{fig:plots_convergence}(right) shows that the accuracy decreases slightly with iterations. We suspect that this loss of accuracy is due to the fact that the updater is only trained on the output of the single-view network, not on its own output. However, training the updater recursively would be more involved. We found that $5$ iterations provide a good trade-off between multi-view coherence and accuracy.

	
\subsection{Robustness}
\begin{figure}[!t]
	\begin{center}
		\includegraphics[width=\linewidth]{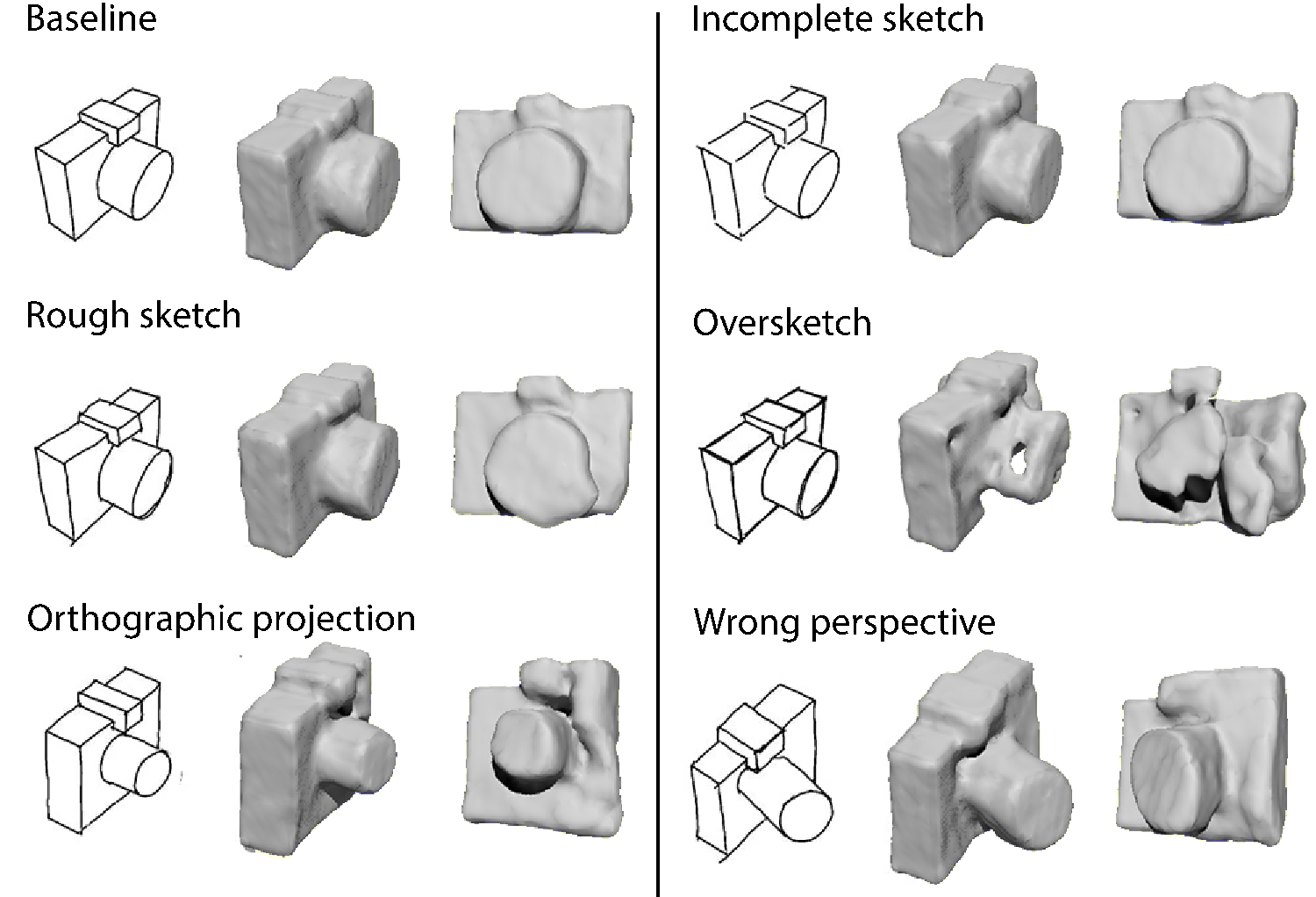}
		\caption{Behavior of the single-view network on various sources of noise. While the network trained on clean drawing tolerates some amount of sketchiness, overshoot and incompleteness, it is sensitive to over-sketching that produces thicker lines than the ones in the training set. Drawing with a very different or wrong perspective yields distorted shapes.}
		\label{fig:robustness}
	\end{center}
\end{figure}

While we trained our networks with clean drawings rendered with perfect perspective, they offer robustness to moderate sources of noise, such as wavy, incomplete or overshot lines and slight perspective distortions, as shown in Figure~\ref{fig:robustness}. However, drawings made under drastically different or wrong perspectives yield distorted shapes. We also observed sensitivity to over-sketching and varying line thickness. An interesting direction for future work would be to render the training data using advanced non-photorealistic rendering, in the hope of achieving invariance to line style.

\subsection{Comparisons}
\begin{figure}[t!]
	\centering
	\includegraphics[width=\linewidth]{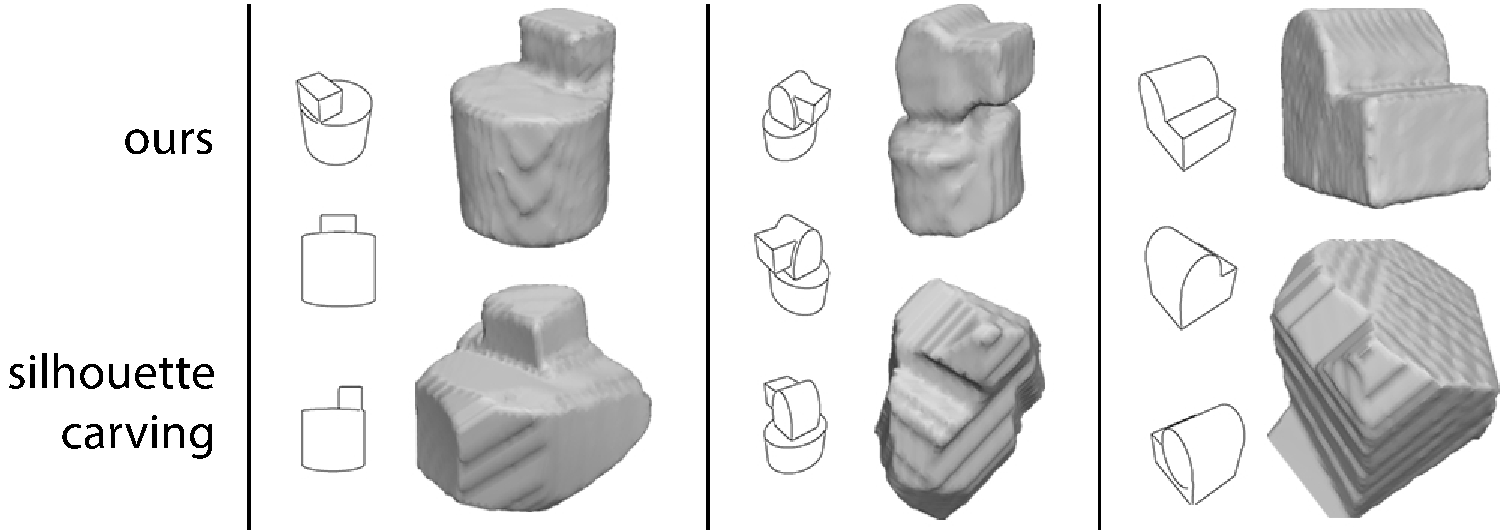}
	\caption{Reconstructed objects using our method (top row) and silhouette carving (bottom row) with 3 random views. Silhouette carving struggles to recover concavities.}
	\label{fig:silhouette_carving}
\end{figure}

\begin{figure}[!t]
		\begin{center}
			\includegraphics[height=0.55\linewidth]{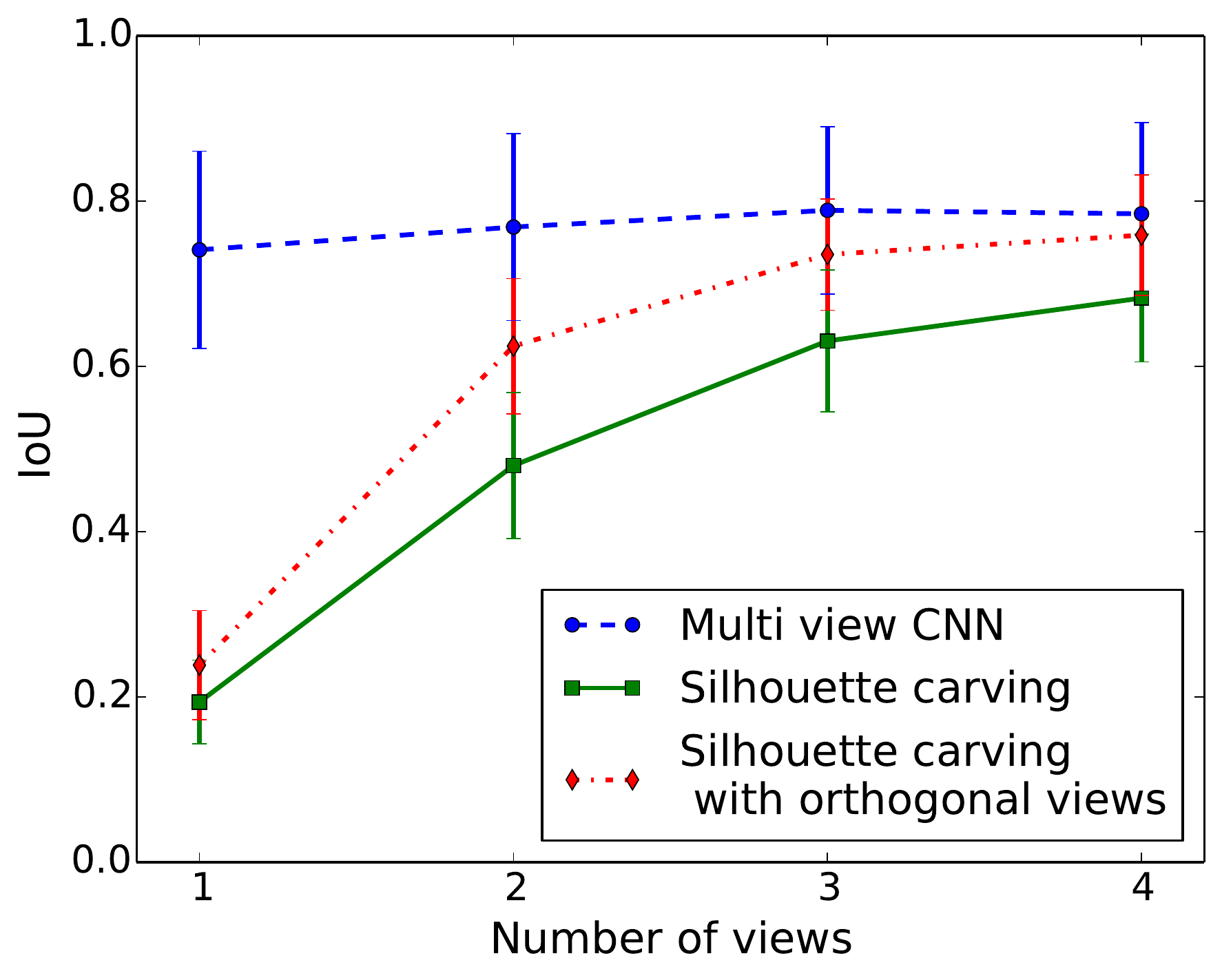}
			\caption{Comparison between our method (blue) and silhouette carving (green and red). The strength of our approach is that it achieves high accuracy from only one view, and remains competitive with silhouette carving with four views. In addition, our method can handle concavities that cannot be recovered by carving.}
			\label{fig:plots_comp_carving}
		\end{center}
	\end{figure}

To the best of our knowledge, our method is the first that can automatically reconstruct a 3D model from a set of multiple perspective bitmap drawings. As a baseline, we compare our approach with a silhouette carving algorithm \cite{Martin1983}. We implemented two versions of silhouette carving for this comparison. The first version takes as input the same drawings as the ones provided to our method, which necessarily includes a 3/4 view for the first drawing to be fed to the single-view network, and different random views for the other drawings. The second version only takes drawings from orthogonal views, which is the most informative setup for silhouette carving. As shown in Figure~\ref{fig:plots_comp_carving}, our approach outperforms silhouette carving in both conditions. In particular, our method achieves a high IoU ratio with as little as one view. Figure~\ref{fig:silhouette_carving} provides a visual comparison between our reconstructions and the ones by silhouette carving. Our approach is especially beneficial in the presence of concavities.

\begin{figure}[!t]
	\begin{center}
		\includegraphics[width=0.36\linewidth]{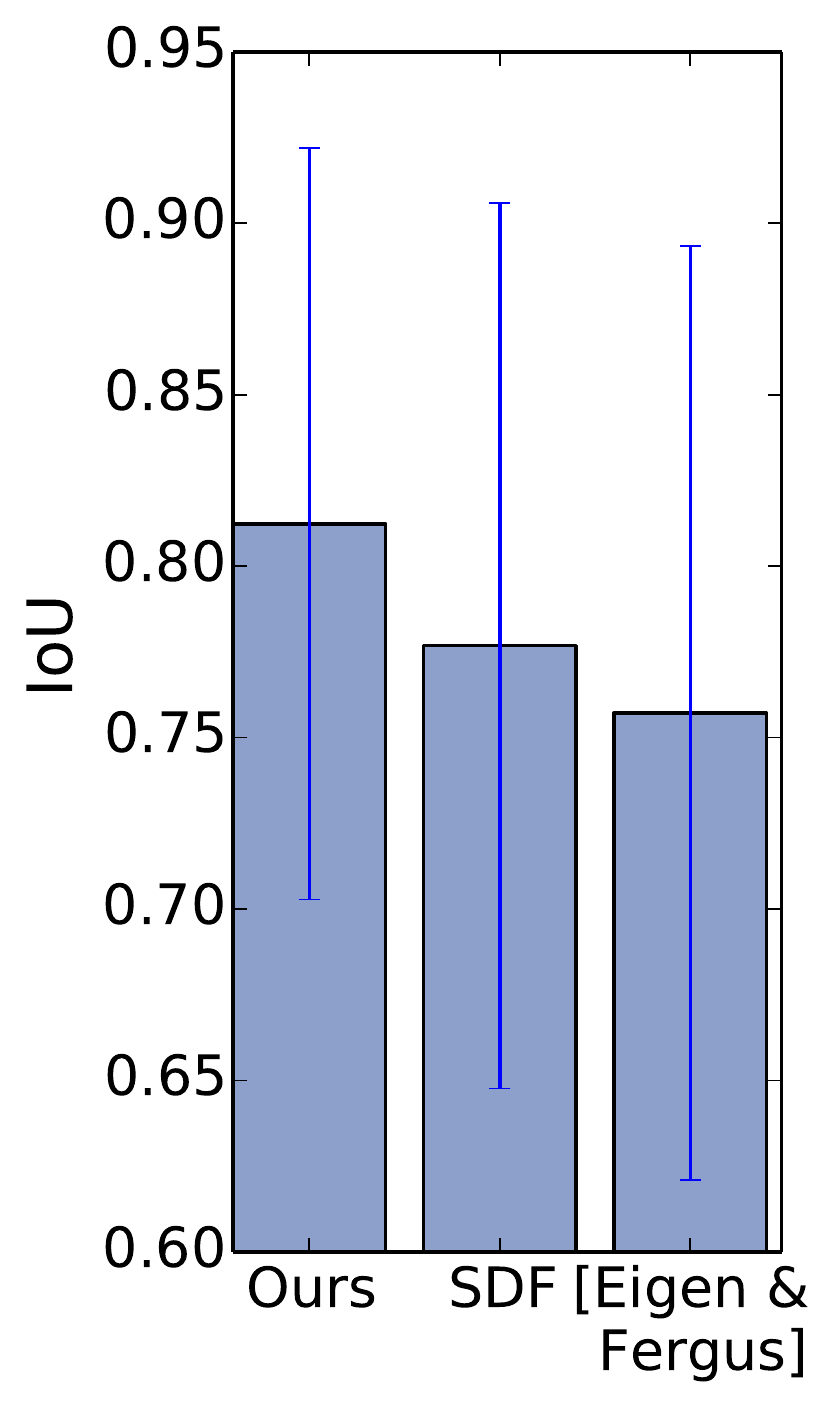}
		\includegraphics[width=0.62\linewidth]{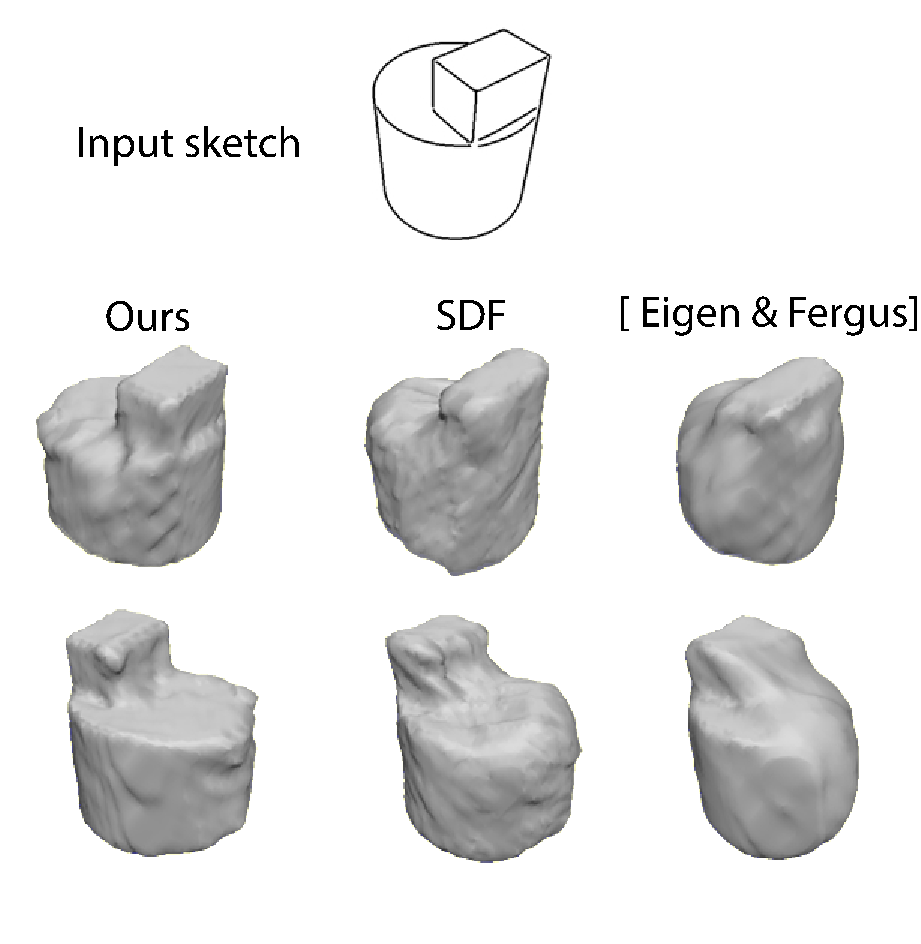}
		\caption{We compare our single-view network with the one of Eigen and Fergus \protect~\cite{eigen2015predicting} and with a network trained to predict a signed-distance function rather than a binary voxel grid. Our design outperforms these two alternatives.}
		\label{fig:comp_diff_nets}
	\end{center}
\end{figure}

Figure~\ref{fig:comp_diff_nets} evaluates our network architecture against several alternative designs. We perform this evaluation on the single-view network since any improvement made on it would directly benefit the updater. A first important design choice to evaluate is the choice of the volumetric representation. While we chose a binary representation of the volume, we also considered a signed distance function. However, our experiments reveal that this alternative representation reduces quality slightly, producing smoother predictions than ours. We also compare our U-net architecture with the multi-scale depth prediction network proposed by Eigen and Fergus~\shortcite{eigen2015predicting}, which we modified to output a multi-channel image. This network follows a similar encoder-decoder strategy as ours but does not include as many skip-connections between multi-scale layers, which also reduces the quality of the prediction.

\subsection{Limitations}
\begin{figure}[!t]
	\begin{center}
		\includegraphics[width=\linewidth]{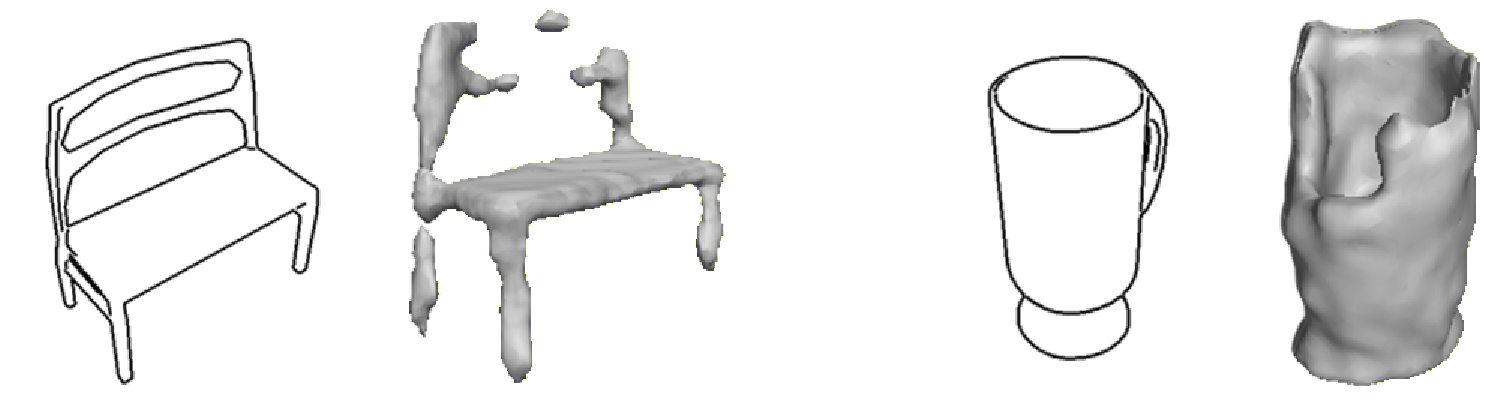}
		\caption{Thin structures are challenging to capture by the $64^3$ voxel grid.}
		\label{fig:thin_structures}
	\end{center}
\end{figure}

Figure \ref{fig:thin_structures} shows drawings with thin structures that are challenging to reconstruct for our current implementation based on a $64^3$ voxel grid. High-resolution volumetric representations is an active topic in deep learning \cite{hspHane17,RieglerUBG17,Fan_cvpr2017} and we hope to benefit from progress in that field in the near future. An alternative approach is to predict multi-view depth maps, as proposed by Lun et al. \shortcite{Lun2017}, although these depth maps need to be registered and fused by an optimization method to produce the final 3D surface.

\begin{figure}[!t]
	\begin{center}
		\includegraphics[width=\linewidth]{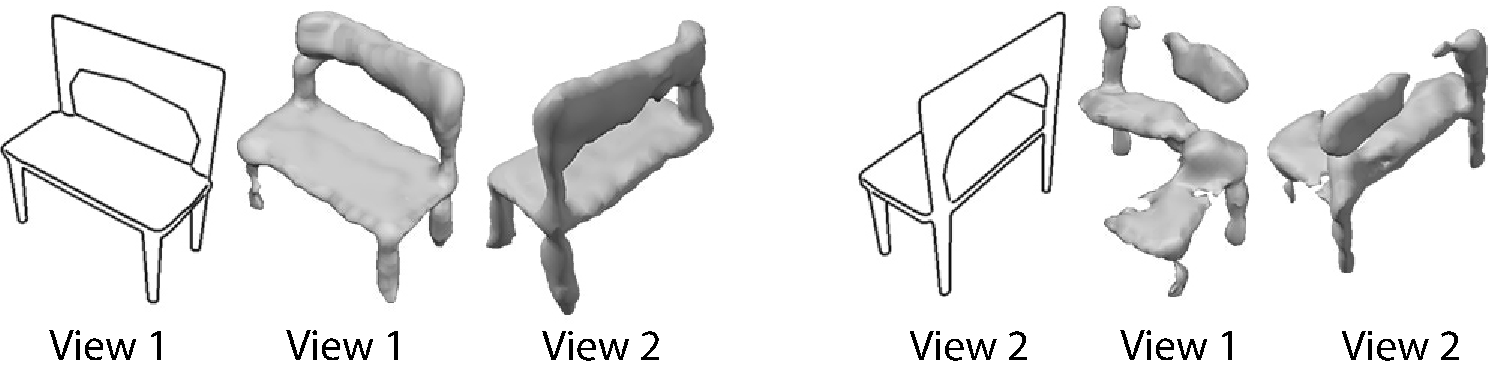}
		\caption{The single-view network performs best on informative viewpoints that minimize occlusions (left). Drawing the same shape from a viewpoint with significant occlusions results in an erroneous prediction (right).}
		\label{fig:ambiguity}
	\end{center}
\end{figure}

Our deep networks also have difficulty interpreting drawings with many occlusions, as shown in Figure \ref{fig:ambiguity}. Fortunately, designers tend to avoid viewpoints with many occlusions since they are not the most informative. Nevertheless, occlusions are inevitable on objects composed of many parts, and we observed that the quality of the reconstruction can reduce as users add more details to their drawings. A simple solution to this limitation would consist in letting the user freeze the reconstruction before adding novel parts. This feature could be implemented by copying the reconstruction in a temporary buffer, and flagging all the lines as construction lines to be ignored by the system. Users could then proceed with drawing new parts which would be interpreted as a new object, and we could display the existing reconstruction and the new parts together by taking the union of their volumes.

\subsection{Performances} 
We implemented our system using the Caffe library for deep learning \cite{jia2014caffe} and OpenGL for real-time rendering in the user interface. Table~\ref{tab:tabTimings} provides timings at test time for an increasing number of views, measured on a desktop computer with an NVidia TitanX GPU, and on a MacBook Pro laptop using only the CPU. Our 3D reconstruction engine scales linearly with the number of views and outputs a prediction in less than a second using GPU and within a few seconds using CPU, on a $64^3$ voxel grid with $5$ iterations of the updater. Our single-view and updater networks occupy $775$MB of memory together.

\begin{table}[!!h]
\centering
\begin{tabular}{  l | c | c | c | c  }
 & 1 view & 2 views & 3 views & 4 views \\
\hline
Desktop GPU (ms) & $140$ & $210$ & $280$ & $350$ \\
Laptop CPU (s) & $1$ & $1.5$ & $2.2$ & $2.9$ \\
\end{tabular}
\caption{Our method scales linearly with the number of input drawings, generating the prediction in less than a second for a $64^3$ voxel grid on a modern GPU.}
\label{tab:tabTimings}
\end{table}

\section{Discussion}
	Research in sketch-based modeling has long been driven by the need for a flexible method capable of reconstructing a large variety of shapes from drawings with minimal user indications. In this paper we explored the use of deep learning to reach this goal and proposed an architecture capable of predicting 3D volumes from a single drawing, as well as fusing information from multiple drawings via iterative updates. This architecture fits naturally in a simple modeling interface allowing users to seamlessly sketch and visualize shapes in 3D. Our approach is modular and we see multiple directions of future research to improve it.
	
	We demonstrated the potential of our system by training it with simple contour drawings. Artists often use other visual cues to depict shape in their drawings, such as hatching to convey shading \cite{hertzmann2000illustrating}, cross-sections to convey curvature directions \cite{shao2012crossshade}, scaffolds and vanishing lines to lay down perspective and bounding volumes \cite{schmidt2009analytic}. An exciting direction of research would be to train our system to generalize to all these drawing techniques. However, achieving this goal may require the design of new non-photorealistic rendering algorithms that formalize and reproduce such techniques \cite{Gori2017}. Going further, style transfer algorithms \cite{Kalogerakis2012} may even enable the synthesis of user-specific training data.
	
	Despite its simplicity, our abstract shape grammar proved sufficient to train our system to reconstruct a variety of man-made objects. We hope that this new application will motivate further research in the design of advanced shape grammars that capture the statistics of real-world objects. 
	
	We used simple thresholding to extract a 3D surface from the predicted occupancy grid. More advanced surface extraction algorithms such as graphcut segmentation \cite{Boykov01} could be used to further regularize the solution, for instance by incorporating a prior on piecewise-smooth surfaces. Alternatively, finer reconstructions may be obtained by training our CNNs with different loss functions. In particular, \emph{adversarial networks} have recently shown an impressive ability to hallucinate fine details in synthesis tasks by combining a generator network with a discriminator that learns to identify if an output is real or synthesized \cite{pix2pix2016}. 
	
	In this work, we explored deep learning as an alternative to hand-crafted geometric optimizations. Fundamentaly, we see these two approaches as complementary and would like to use our predictions to initialize precise constraint-based optimizations that would strictly enforce regularities such as parallelism, orthogonality and symmetry \cite{li_globFit_sigg11,Xu2014}.
	
	Finally, while we developed our iterative updater architecture to reconstruct objects from drawings, a similar architecture could be used for multiview 3D reconstruction from photographs given calibrated cameras. The challenge of such an approach is to obtain a sufficiently large amount of training data that covers not only different shapes, but also different textures, materials and lighting conditions as encountered in realistic scenes.

\begin{acks}
Many thanks to Yulia Gryaditskaya for sketching several of our results, and for her help on the renderings and video. This work was supported in part by the ERC starting grant D$^3$ (ERC-2016-STG 714221), the Intel/NSF VEC award IIS-1539099, FBF grant 2018-0017, ANR project EnHerit (ANR-17-CE23-0008),
research and software donations from Adobe, and by hardware donations from NVIDIA.
\end{acks}

\bibliographystyle{ACM-Reference-Format}
\bibliography{deep_sketch}

\newpage
\section*{Appendix}

We adapt our architecture from \cite{pix2pix2016} by reducing the number of layers of the decoder part.

Let $\mathsf{(De)C}k$  denote a (De)Convolution-BatchNorm-ReLU layer with $k$ filters (output has $k$ channels). 
$\mathsf{(De)CD}k$  denotes a (De)Convolution-BatchNorm-Dropout-ReLU layer with a dropout rate of 50\%. 
All convolutions are 4×4 spatial filters applied with stride 2. Convolutions in the encoder downsample by a factor of 2, whereas deconvolutions in the decoder upsample by a factor of 2. 

The encoder-decoder architecture consists of: 
 \begin{center}
 	\includegraphics[width=1.1\linewidth]{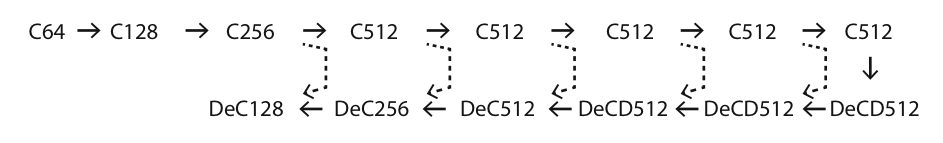}
 \end{center}

After the last layer of the decoder, a SoftMax is applied followed by a classification loss (multinomial logistic loss). We then keep only one channel over two (the one containing the probability of occupancy) to get a voxel grid of dimension 64.

As an exception to the above notation, Batch-Norm is not applied to the first $\mathsf{C64}$ layer in the encoder. All ReLUs in the encoder are leaky, with slope 0.2, while ReLUs in the decoder are not leaky. 

The skip connections (shown as dashed arrows ) consist in concatenating the output of a convolution in the encoder to a deconvolution of the same size in the  decoder.

\end{document}